\documentclass{aa}  

\usepackage{subcaption}
\usepackage{graphicx}
\usepackage{ragged2e}
\usepackage{caption}
\usepackage{amsmath}
\usepackage{booktabs}
\usepackage{float}
\usepackage{placeins}
\usepackage{orcidlink}

\def\kms{km~s$^{-1}$\/}

\usepackage{txfonts}
\usepackage{natbib}
\bibpunct{(}{)}{;}{a}{}{,}
\DeclareUnicodeCharacter{2033}{''}
\begin{document}

\title{A Broker Integrated Algorithm for Gravitational Wave  –  Electromagnetic Counterpart Searches in O4a and O4b Runs}

\titlerunning{Broker integrated GW – EM counterpart search (O4a–O4b)}
\authorrunning{Bommireddy et al.}

\author{ 
  Hemanth Bommireddy\inst{1,2,3,4} \orcidlink{0009-0007-4271-6444} \and
  Francisco Förster\inst{1,2,3,4,5} \orcidlink{0000-0003-3459-2270} \and
  Isaac McMahon\inst{6} \orcidlink{0000-0002-4529-1505} \and
  Manuel Pavez-Herrera\inst{7} \orcidlink{0009-0000-7125-9966} \and
  Regis Cartier \inst{8} \orcidlink{0000-0003-4553-4033} \and
  Felipe Olivares~E.\inst{9} \orcidlink{0000-0002-5115-6377} \and
  Lorena Hernández-García\inst{10,11,4} \orcidlink{0000-0002-8606-6961}\and
  Mary Loli Martínez-Aldama\inst{4,12,13}\orcidlink{0000-0002-7843-7689}  \and
  Alejandra Mu\~noz Arancibia \inst{4,5}\orcidlink{0000-0002-8722-516X}
}

\institute{
   Departamento de Astronomia (DAS), Universidad de Chile, Camino el Observatorio 1515, Las Condes, Santiago, Chile 
   \email{hemanth@das.uchile.cl}\and
  Data and Artificial Intelligence Initiative (IDIA), Faculty of Physical and Mathematical Sciences, Universidad de Chile, Santiago, Chile \and
  Centro de Excelencia en Astrofisica y Technologia Afines, Av. Vicuña Mackenna 4860, San Joaquín, Santiago, Chile\and
  Millennium Institute of Astrophysics, Nuncio Monseñor Sotero Sanz 100, Of. 104, Providencia, Santiago, Chile \and
  Center for Mathematical Modeling, Universidad de Chile, Beauchef 851, Santiago 8370456, Chile \and
  Physik-Institut, Universität Zürich, Winterthurerstrasse 190, CH-8057 Zürich, Schweiz
  \and Instituto de Astrofísica, Facultad de Física, Pontificia Universidad Católica de Chile, Campus San Joaquín, Av. Vicuña Mackenna 4860, Macul Santiago, Chile, 7820436
  \and Centro de Astronomia, Universidad de Antofogasta,  02800, Antofagasta, Chile
  \and UKIRT Observatory, Institute for Astronomy, 640 N.\ A'ohoku Place, University Park, Hilo, Hawai'i 96720, USA
  \and Instituto de Estudios Astrof\'isicos, Facultad de Ingenier\'ia y Ciencias, Universidad Diego Portales, Av. Ej\'ercito Libertador 441, Santiago, Chile
  \and Centro Interdisciplinario de Data Science, Facultad de Ingenier\'ia y Ciencias, Universidad Diego Portales, Av. Ej\'ercito Libertador 441, Santiago, Chile
  \and Astronomy Department, Universidad de Concepción, Casilla 160-C,  4030000, Concepción, Chile
  \and Millennium Nucleus on Transversal Research and Technology to Explore Supermassive Black Holes (TITANs), Chile
}

   %\date{Received September 15, 2025; accepted %March 16, 2026}

% \abstract{}{}{}{}{} 
% 5 {} token are mandatory
 
  \abstract
  % context heading (optional)
  % {} leave it empty if necessary  
   {We present an automated framework to search for optical counterparts of LIGO–Virgo–KAGRA (LVK) gravitational-wave (GW) superevents using public Zwicky Transient Facility (ZTF) alerts processed through the ALeRCE broker.}
  % aims heading (mandatory)
   {The goal is to filter and identify optical transients potentially associated with binary black hole (BBH) mergers during the LVK O4a and O4b observing runs.}
  % methods heading (mandatory)
   {Using the automatic learning for rapid classification of events (ALeRCE) infrastructure, we spatially query ZTF alerts within GW localization regions and apply machine-learning classifiers, host-galaxy crossmatching, and temporal cuts within 200 days post-merger to isolate plausible candidates.}
  % conclusions heading (optional), leave it empty if necessary 
   {Our search yielded one candidate in O4a and four in O4b, several consistent with the supernova–TDE regime. This work demonstrates that public alert brokers can set a  baseline to perform systematic searches for electromagnetic counterparts to GW superevents in current and future observing runs.}
   {Our algorithm provides a way to search for the BBH counterparts for all significant LVK GW superevents using survey telescope alerts. The search along with the accompanying analysis demonstrates the significance of the counterparts candidates, with one candidate ultimately identified as a transient event consistent with a Bowen fluorescence flare in an (now discarded) active galactic nucleus (AGN).}


   \maketitle
%
%-------------------------------------------------------------------
\def\ndet{391}
\section{Introduction} \label{Introduction}
   Since the first observation of a gravitational wave (GW) by the LIGO–Virgo–KAGRA (LVK) detectors \citep{abbott.2016}, a total of 391 gravitational wave candidates have been detected, as of the end of the fourth observing run (O4). Of these events, binary black hole (BBH) mergers are the dominant class, accounting for more than 98\% of all the detections. Broadly, the formation channels of the mergers can be classified into isolated and dynamic \citep{Mapelli_2020}. The  BBH event GW150921 has been proposed to be an example of the dynamic formation channel, with a possible association with an active galactic nucleus (AGN) disk environment \citep{Graham_2020}, and a light curve consistent with a post-merger disk flaring event (object ZTF19abanrhr).

    Several studies  have estimated BBH merger rates in AGN accretion disks \citep{Ford_2022,Gr_bner_2020,mckernan.2018}. Recently, \cite{Ford_2022} have estimated the rate of $R \sim 24\  \mathrm{Gpc}^{-3}\mathrm{yr}^{-1}$, analytically calculated based on binary lifetime, binary fraction, active life time of galactic nuclei, and their AGN fraction. This rate broadly falls in the range of rates predicted by the LVK collaboration $R \sim (17.9-44) \ \mathrm{Gpc^{-3}} \mathrm{yr^{-1}}$ for $z=0.2$ \citep{abbott.2023}, scaling with redshift as $(1+z)^{k}$, with $k=2.9^{+1.7}_{-1.8}$ for $z\leq 1$. In the AGN environment, high escape velocities allow retention of BBH merger products \citep{Samsing_2022} and its dissipative gas environment allows BHs to pair up again and grow up through hierarchical mergers \citep{Tagawa.2020, Tagawa.2021, Gilbaum.2025}.
    The interaction of the merged BBH with the AGN accretion disk can produce an electromagntic (EM) counterpart. The merged product receives a recoil velocity, which is a function of binary mass asymmetry and spin orientation \citep{Baker.2008}. The remnant might interact with the gas via Bondi-Hoyle-Littleton (BHL) accretion. This can lead to the production of detectable luminosity at super Eddington rates \citep{Graham_2020}, along with the jetted outflows that allow radiation to escape and emerge \citep{wang.2021b}. The counterpart could emerge weeks to months after the event occurs, modulated by the AGN disk diffusion timescale. Modern time domain surveys scanning the sky with weeks to months cadence could be helpful in identifying and monitoring these transients.
    
Challenges lie in distinguishing these counterparts from innate AGNs variability, and different types of nuclear transient events such as tidal disruption events (TDEs), supernovae (SNe), microlensing close to the super massive black hole (SMBH). Moreover, alternative mechanisms have been developed  that could result in the production of an EM counterpart along with the GW (\citealt{McKernan2012,Bartos2017,Kimura.2021,Perna2021, wang.2021b, rodríguezramírez2025opticaluvflaresbinary, Tagawa_2023}).

   Many collaborations such as GW-MMADS \citep{keerthi.2024}, GRANDMA \citep{grandma.2022} and DESGW \citep{desgw.2020}, are actively searching for the counterparts of gravitational sources. Web-based target and observation managers from centralized portals such as SAGUARO \citep{Hoss.2024}, and SkyPortal \citep{coughlin.2023} have been built to reduce the need for human intervention and improve coordination to search for the counterparts. However, this search is usually done only when the GW event localization area is quite small ($<100\ \mathrm{deg}^2$). 
   
   In the era of big data astronomy, `alert brokers' ingest, annotate, filter and classify astronomical alerts from survey telescopes such as Zwicky transient Facility \citep[ZTF,][]{Bellm_2018},
   Asteroid Terrestrial-impact Last Alert System  \citep[ATLAS,][]{Denneau.2014}, La Silla Schmidt Southern Survey \citep[LS4,][]{Adam.2025}, and Legacy Survey of Space and Time \citep[LSST,][]{Ivezic.2018} for to astronomy community. Among the existing alert brokers, ANTARES \citep{Matheson.2021}, and FINK \citep{Moller.2021} also crossmatch alerts with LVK sky maps.

   %Among the AGN flares, Bowen Fluorescence Flares (BFFs) are of a unique kind \citep{wiseman.2025,Trakhtenbrot_2019}. They could result as consequence of tidal disruption of a star with the central SMBH or intense accretion of dust at the SMBH. BFFs spectra typically exhibit prominent NIII $\lambda4640$ and OIII $\lambda 3133$ emission features, which are uncommon in typical AGN spectra \citep{vander.2001}. The presence of strong Bowen fluorescence lines serves as a diagnostic of intense ultraviolet (EUV; down to $\sim100 \ Å$) radiation. Their observed broad profiles suggest an origin within the AGN broad-line region (BLR). An accretion-driven mechanism could naturally account for both the emergence of the Bowen lines and the high optical luminosities observed in these events. A few  have been reported in the transient name server \citep[TNS,][]{Gal.2021}, namely AT2025cfyh, AT2023zgo, AT2021loi, AT2020afhd, and AT2017bgt. 
   %In our work, instead of a targeted search, we take advantage of the public alerts from the Zwicky Transient Facility (ZTF), in tandem with the Machine Learning (ML) based infrastructure of Automatic Learning for Rapid Classification of Events broker (ALeRCE; Forster et al. 2021) used for transients classification, with other data science tools to search for the counterparts of BBH mergers in AGN.

   In our work, we take advantage of the public alerts from ZTF and classification algorithms of the automatic learning for rapid classification of events \citep[ALeRCE,][]{forster2021} broker, to search for the counterparts of BBH mergers in AGN in an untargeted fashion, i.e., considering all LVK superevents and their crossmatches. 
   
   We introduce an algorithm that utilizes the ALeRCE services to search for the optical counterparts of GW sources during the O4 run. Given the volumes of data generated by ZTF in the form of alerts, we built an automated pipeline, which on receiving LIGO alerts (effectively, a sky map), filters only those ZTF events whose localization and redshift (inferred from the LVK alert) are consistent with that of AGN with detectable activity within 200 days post-merger.

   This paper is organized as follows. In Section \ref{Data} we introduce the data and the ALeRCE broker. In section \ref{methodology} we describe the algorithm designed to filter the possible counterparts from the alerts. In Section \ref{results}, we explain the results of the application of the algorithm on the LVK O4a and O4b runs, including the most interesting candidates, and in Section \ref{analysis} we analyze the results, comparing the expected number and properties of associated flares with our previous findings. Finally, our discussions and conclusions are summarized in Sections \ref{discussion} and \ref{conclusions}.

%--------------------------------------------------------------------
\section{Data} \label{Data}

%-------------------------------------- Two column figure (place early!)
\subsection{ZTF}
 
    The ZTF \citep{Bellm_2018,bellm.2019} has been surveying the northern sky every $\sim$ 3 days in the g, r and i optical filters since 2018 and offers different services such as: (1) a public alerts system for real time domain science; (2) data releases (DRs) every two months, including photometric measurements on the science images; and (3) a Forced Photometry service on demand per source, including the photometry measurements on the reference-subtracted science images. \\
    For the alert to be generated, a source has to show a variation at least five times above the background noise of the difference image (5$\sigma$).

\subsubsection{ZTF Forced Photometry} 

We retrieved data from the ZTF Forced Photometry Service \citep{masci2019, masci2023}. The measurements from this service are based on reference-subtracted images, isolating the variable nuclear component and reducing contamination from host-galaxy emission compared to public Data Releases. The light curves were constructed by removing measurements with bad-quality flags, requiring seeing $< 5''$ and positive flux uncertainties. Relaxed cuts were applied to preserve faint nuclear variability potentially associated with GW events. No color correction was applied. The difference PSF magnitudes were converted into apparent magnitudes following \citet{forster2021}.

 \subsection{ALeRCE annotations}
 ALeRCE is a Chilean-led broker that is processing the alert stream from ZTF, that was selected as a Community Broker for the Vera C. Rubin Observatory and its Legacy Survey of Space and Time (LSST), and that aims to include other large etendue survey telescopes.

 ALeRCE \footnote{\url{https://github.com/alercebroker}} uses a pipeline that includes real-time ingestion, aggregation, cross-matching, machine learning (ML) classification, and visualization of the ZTF alert stream. We use two classifiers: a stamp-based classifier \citep{carrasco.2021}, that uses image stamps as input and is designed for rapid classification; and a light curve-based classifier \citep{Sanchez_Saez_2021}, that uses the multiband flux evolution to achieve a more refined classification. 
 
 All alerts associated with new objects undergo the stamp-based classification, that provides a quick classification into a five-class taxonomy (supernova, active galactic nuclei, variable star, asteroid, and bogus). The stamp classifier consists of a rotationally invariant convolutional neural network which uses information from the image stamps and the alert’s metadata.

\subsection{GW alerts/skymaps}
GW alerts are distributed in JSON and AVRO formats, containing key metadata such as the estimated luminosity distance, merger classification (e.g., binary black hole, binary neutron star), and significance of the detection. All publicly released alerts are archived in the Gravitational-Wave Candidate Event Database (GraceDB)\footnote{\url{https://gracedb.ligo.org/superevents/public/O4/}}. For this work, we retrieve the alerts and associated data products after public release from GraceDB, rather than processing them in real time. Each alert includes a URL to the corresponding localization skymap (typically a \texttt{.fits.gz} file), which we use to construct probability maps and perform spatial queries for potential electromagnetic counterparts, as described in Section~\ref{methodology}. The 3D localization skymaps are subject to revision when the final O4 catalog is released, and modest changes in the sky-localization volumes or distance posteriors may affect some of the derived association probabilities.

\section{Methodology} \label{methodology}

\subsection{Retrieving GW Skymaps}

\begin{figure}[h!]
\includegraphics[width=9cm]{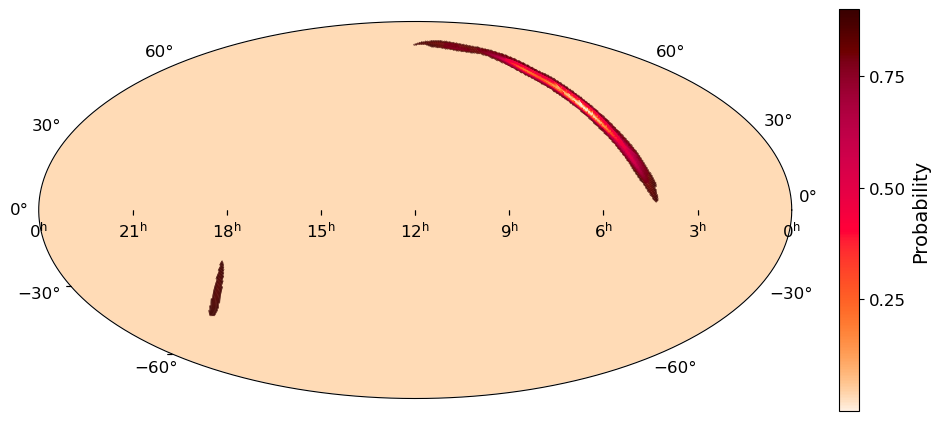}
\centering
\caption{LIGO skymap of the GW event S240902bq with color scale representing the probability contours.}
\label{fig:skymap}
\end{figure}

The first step to search for the optical counterparts of LIGO/VIRGO/Kagra (LVK) superevents is to obtain their projected spatial probability distribution, which will be referred to as skymaps. LVK  distribute these skymaps in standard HEALPix \citep{Gorksi.2005} representation in flat and multi-resolution formats. We use the multi-resolution skymaps for the counterparts search, assuming that the LVK superevents are within the 90\% integrated probability contours.
In general, these skymaps can  be obtained from the following sources:
\begin{enumerate}
    \item the public General Coordinates Network (GCN) alert stream.
    \item the Gravitational-Wave candidates event database (GraceDB) web interface. \footnote{\url{https://gracedb.ligo.org/}}
    \item the gracedb-sdk client \footnote{\url{https://github.com/lpsinger/observing-scenarios-simulations/blob/main/scripts/get-public-alerts.py}}.
\end{enumerate}
The skymaps in this study were built following the instructions in the emfollow website, a LVK public alerts user guide\footnote{\url{https://emfollow.docs.ligo.org/userguide/tutorial/multiorder_skymaps.html}}.
An example skymap is depicted in 
 figure \ref{fig:skymap}.

\subsection{Performing spatial queries within skymaps}

Once the skymaps are available, we compare them with ZTF optical-based alerts. We use the ALeRCE broker to access the ZTF alert data.

The time and location of every ZTF alert required is available in ALeRCE PostgreSQL database. The spatial query is performed using the Quad Tree Cube \citep[Q3C,][]{2006ASPC..351..735K} PostgreSQL extension. In Q3C, spatial indexing is achieved by projecting the sides of an imaginary cube into the celestial sphere, where each side is divided into multiple quad tree structures for more efficient querying. In this implementation, the query region is defined by the vertices of a polygon, in this case, the 90\% probability contour of the LVK skymaps. However, there are two limitations associated with using Q3C: (1) query polygons must not exceed 25 degrees in diameter and (2) the number of vertices that define the polygon must be fewer than 100.

.

\begin{figure}[h!]
\includegraphics[width=8cm]{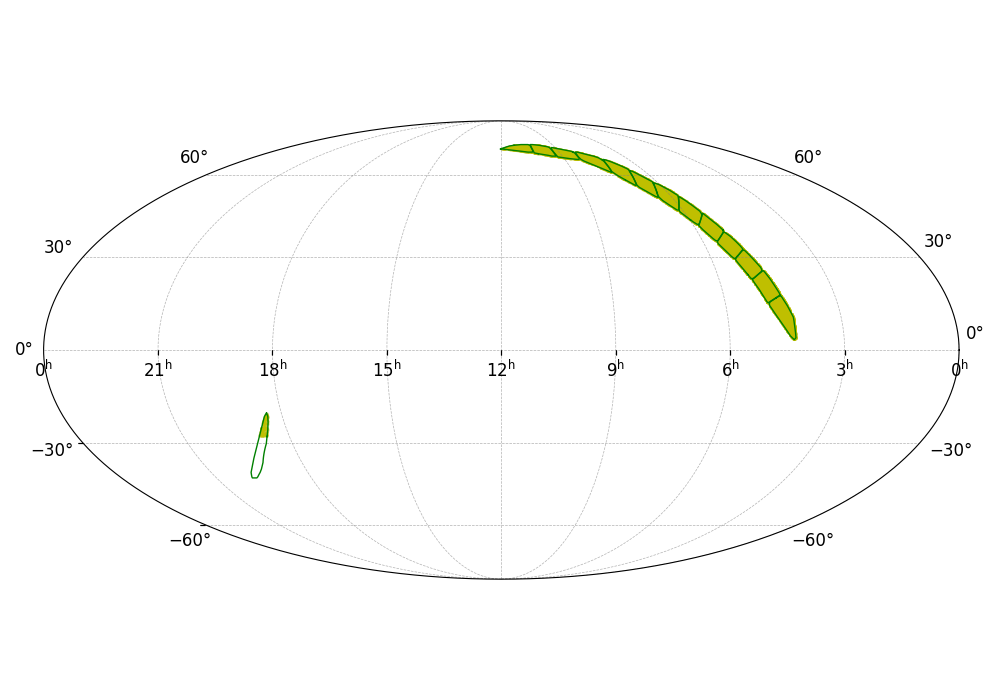}
\centering
\caption{Skymap converted to bounded polygons (green boundaries) filled with ZTF pubic alerts in yellow }
\label{fig:bound}
\end{figure}

Given the limitations and the need to efficiently search within large GW localization regions, we begin by transforming the skymap pixels into a cloud of points. Then, we partition the points into multiple spatial clusters using the K-Means algorithm. Each cluster consists of a finite set of HEALPix pixels, each containing both spatial and probability information. To enable spatial querying, we convert the pixels within each cluster into bounding polygons using the \texttt{alphashape} library. In contrast to convex hulls, these polygons need to adopt a concave hull geometry as they more accurately follow the contours of the skymap. The degree of concavity is controlled by the alpha parameter. For our work, we found that an alpha value of 0.2 achieves an effective balance between contour fidelity and overfitting. These resulting polygons are then used as spatial regions for querying transient candidates, as illustrated in Figure \ref{fig:bound}.

The optical counterparts may be long-lived due to the long diffusion timescales of the supermassive black hole accretion disk as BHL accretion takes place. Following \citep{Graham_2023}, we define a maximum search timescale of 200 days post merger, to be sensitive enough to kick velocities $\geq 1 \mathrm{km}\mathrm{s^{-1}}$.

\subsection{Refining Candidates}
Having obtained the database with all the ZTF public alerts (hereafter alerts) within the LVK skymap, we built an algorithm to narrow down the search for the counterparts as described below.

\begin{enumerate}
    \item A typical spatial query of the alerts in a GW skymap contains $\sim 10^6$ objects. As an initial filtering step, we select the alerts in proximity to known AGNs using the MILLIQUAS catalog  \citep[hereafter MQUAS,][]{flesch.23}, a compilation of approximately one million quasars with associated redshifts and other properties. Given the resolution of ZTF, we retained only those events located within 3″ of MQUAS AGNs. This catalog was preferred due to its inclusion of redshift information, which is crucial for distance-based filtering. However, it is important to note that MQUAS predominantly contains sources from the northern hemisphere, a bias stemming from the sky coverage of contributing AGN surveys. Moreover, the sample has contamination from other sources, which necessitates the use of additional filtering based on ALeRCE features.

\item As a secondary filtering step to exclude likely bogus detections, variable stars, and asteroids, we apply the ALeRCE classifiers \citep{carrasco.2021,paula.2021}. The stamp classifier utilizes the first image cutouts associated with each alert; specifically the science, reference, and difference images along with Pan-STARRS metadata to assign probabilistic classifications. These include supernova (SN), active galactic nucleus (AGN), variable star (VS), asteroid, and bogus categories. We retain only those candidates for which the combined probability of the SN and AGN classes is greater than or equal to 0.5.

\item 
We also employ the light curve classifier, which uses variability features derived from ZTF photometry and color information from both ZTF and ALLWISE \citep{cutri.2014}. This classifier implements a balanced random forest model trained on sources with at least $n_{\mathrm{det}} \geq 6$ detections in either the $g$ or $r$ band. Candidates are retained only if the total probability assigned to the SN or AGN subclasses is greater than or equal to 0.5.

\item An additional filter is applied to exclude star-like sources, incorporating data from the Pan-STARRS  \citep[PS1,][]{chambers.2016} survey. This step keeps candidates with a ZTF real/bogus score $\geq 0.5$ and a star-galaxy score (sgscore) $\leq 0.5$. Note that the sgscore varies between 0 and 1 for galaxies and stars, respectively. This ensures that only real events of extragalactic origin are considered.

\item The fourth cut is based on the location of the event with respect to the Galactic and solar system planes. First, we keep alerts that occur in regions with a low Galactic attenuation ($A_{\nu}< 1$). For events in proximity to the solar system plane ($<20 \ \mathrm{deg}$), we require multiple detections ($\mathrm{ndet}\  >1$).

\item The fifth and final cut is based on the reported luminosity distance from the LVK events. Only candidates whose MQUAS reported distance is within 2-sigma of the LVK luminosity distance mean are retained.

The source code details are provided in the \textit{Data Availability} section below.

\end{enumerate}

\subsection{Effectiveness of the stamp classifier}

\begin{figure}[h!]
\includegraphics[width=8cm]{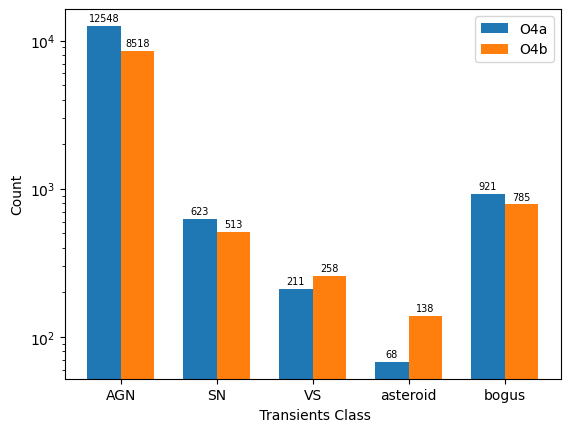}
\centering
\caption{Distribution of classes by stamp classifier, for the events originating with 1.4" of MQUAS AGN catalog for O4a and O4b runs. }
\label{fig:dist1}
\end{figure}

As discussed earlier, the impurity of the MQUAS catalog necessitates the use of ALeRCE classifiers. To evaluate the impact of the stamp classifier filter within our algorithm, we categorized the alerts spatially matched to MQUAS AGNs (within 1.4″) using the stamp classifier for both the LVK data. This spatial crossmatch yielded approximately $\sim 15,059$ and $\sim 10,790$ events for O4a and O4b, respectively. Upon applying the ALeRCE stamp classifier, $\sim 12,544$ and $\sim 8,495$ of these events were classified as AGN, with other classes occurring in smaller numbers, as shown in Figure~\ref{fig:dist1}. Even though the majority of spatially matched events are already classified as AGN, this filter provides a cleaner version of the catalog.

\subsection{ Predicting Chirp Mass} \label{mass}
For GW superevents which are released to the public during an ongoing LVK observing run, information about the masses involved in the merger is not publicly available immediately\footnote{Chirp mass information is being circulated in the GCN circulars from June, 2025 in the form of probability distribution among the mass range bins}. However, for time-critical counterpart follow-up of GW events, a rough estimate of the chirp mass of the system can be estimated using the publicly available localization skymap and machine learning techniques. This is possible because the chirp mass scales directly with the amplitude and luminosity distance of the gravitational wave signal. The luminosity distance is given in the skymap, and information on the Signal-to-Noise Ratio (S/N), and amplitude by proxy, is encoded into the size of the uncertainty region in both luminosity distance and sky area.

We use an Ordinal Class Probability Density Function (OCP) algorithm, originally developed for photometric redshift approximation, but here adapted for mass approximation \citep{rau.2015}. We define classes using linearly equally divided bins on the chirp mass range $[0, 75]\;M_{\odot}$, covering the compact object mergers observed by LVK \citep{lvk.2025} and predicted by population synthesis models \citep{bel.2020, gerosa.2021}. We train a Random Forest Classifier, provided by the \textsc{sklearn} package \citep{sklearn.2011}, to estimate the probability that a given event has a chirp mass inside each mass bin. We then use a Gaussian Kernel Density Estimation (KDE) to estimate a continuous probability density function, and from this KDE we obtain a final estimated median value of the chirp mass with upper and lower bounds. The input vector for each event is given by the HEALPix pixel index, probability, luminosity distance, and luminosity distance error for each of the 20 highest probability pixels in the event skymap, downsampled to HEALPix NSIDE=8. We find that these values are sufficient to encode the direction, amplitude, and distance for chirp mass estimation.

For training data, we simulated 368,000 events using BAYESTAR \citep{Singer_2016}. We used a logarithmic component mass distribution on the range $[1.2, 75]\;M_{\odot}$ and a distance distribution which was uniform in volume out to 6000 Mpc. We assumed aligned spins from the range $[0, 0.05]$ and a network S/N detection threshold of 8. We used the default noise curves for the LIGO/Virgo detectors in O4 implemented in BAYESTAR. Out of our simulated events, 13,757 events surpassed the network detection threshold and were used for training. We find that we generally recover true chirp mass values with $~50\%$ error bounds and a slight bias at the mass range boundaries (see figure \ref{fig:algper}).

\section{Results} \label{results}
During the O4 observing run of the LVK Collaboration, significant improvements were implemented in the low-latency detection pipelines, particularly in background estimation and event ranking, allowing for slightly lower false-alarm-rate thresholds. Combined with enhanced detector sensitivities relative to previous runs, these updates resulted in a substantial increase in the number of detected low-significance candidates, including several high–skymap-area, single-detector triggers \citep{lvk.2025, Petrov_2022, Soni_2025}. Single-detector superevents generally have poor sky localization and low astrophysical probability, making them less suitable for electromagnetic follow-up. Therefore, in this work, we consider only multi-detector detections, which provide higher network S/N and more reliable localization posteriors for crossmatching with optical transient surveys and AGN catalogs.

\begin{figure}[H]
    \centering
    \begin{subfigure}{0.39\textwidth}
        \includegraphics[width=\linewidth]{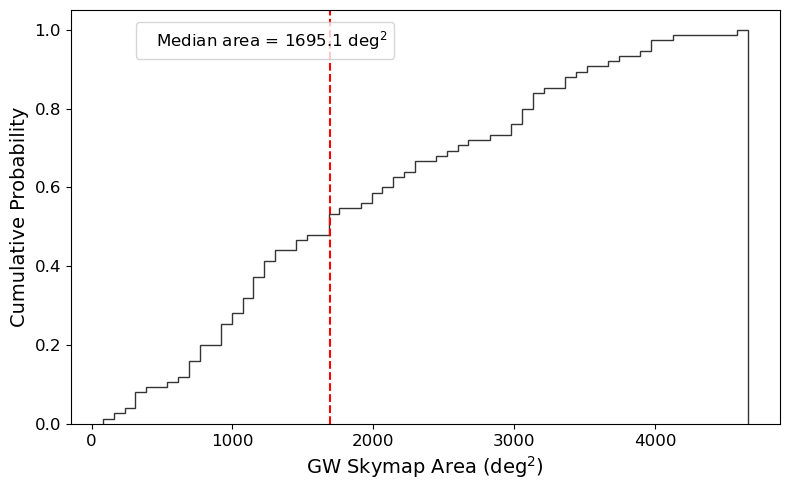}
        \caption{O4a}
        %\label{fig:ztf_a}
    \end{subfigure}
    \hfill
    \begin{subfigure}{0.39\textwidth}
        \includegraphics[width=\linewidth]{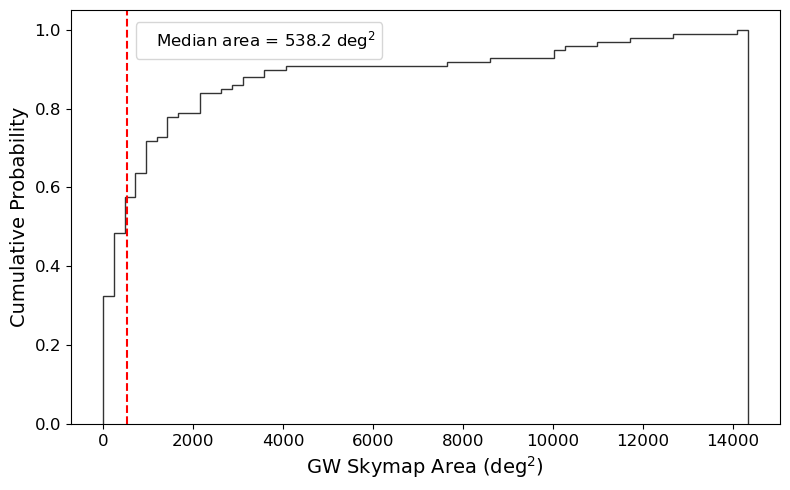}
        \caption{O4b}
        %\label{fig:ztf_b}
    \end{subfigure}
    \caption{Cumulative distribution function (CDF) of the skymap areas for O4a (top) and O4b(bottom) with median area values (vertical red lines)}
    \label{fig:medarea}
\end{figure}

The distribution of areas and the median area for O4a and O4b runs is shown in the Figure \ref{fig:medarea}. Involvement of additional detector during O4b run, demonstrated the decrease in median area from O4a to O4b run.

\begin{figure}
\includegraphics[width=8cm]{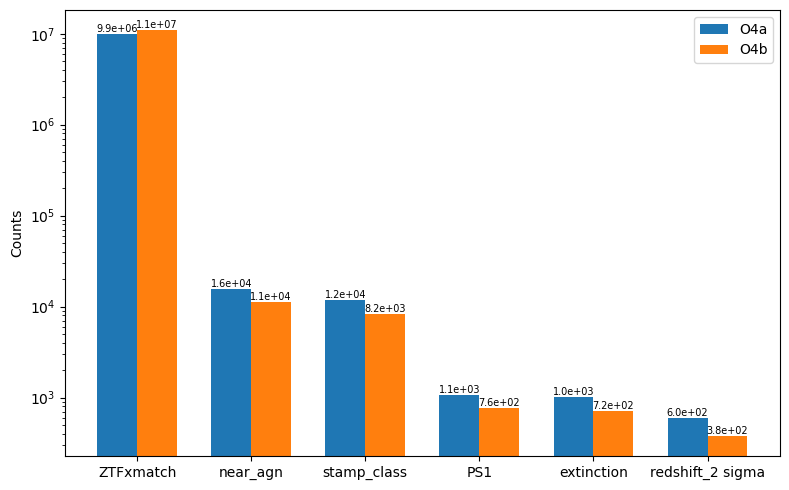}
\centering
\caption{Total number of events for both runs including all filters}
\label{fig:o4aztf}
\end{figure}

\begin{figure}
\includegraphics[width=8cm]{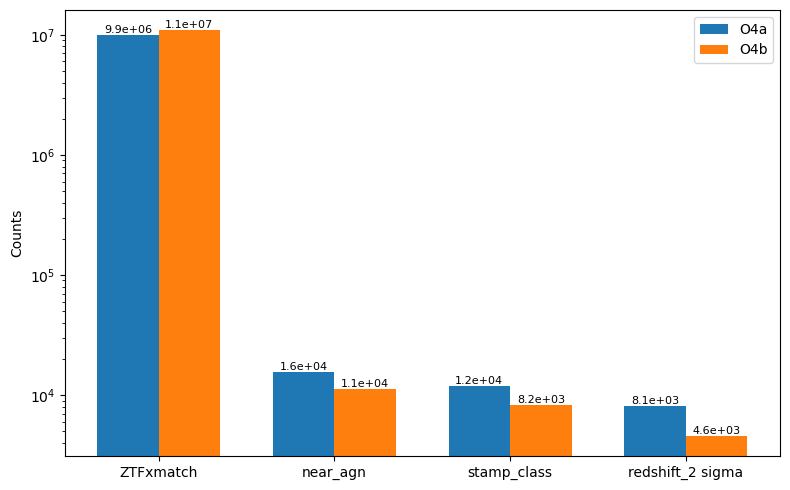}
\centering
\caption{Total number of events for both runs, excluding PS1 and location based filter}
\label{fig:o4bztf}
\end{figure}

Figures \ref{fig:o4aztf} and \ref{fig:o4bztf} represent the number of events through different filters/cuts applied on alerts in the O4a and O4b runs. A noticeable feature is that the number of alerts in the proximity of MQUAS AGNs differ by a factor of ~1000 to that of the total ZTF-LVK crossmatched events.

\subsection{ZTF Candidates}
Implementing the algorithm on multi-detector gravitational-wave superevents yielded 602 and 380 candidates for the O4a and O4b observing runs, respectively. After visual inspection and excluding alerts already classified as SN or TDE on the Transient Name Server (TNS), we identified one interesting candidate, ZTF23abqkwzr, from O4a, and four interesting candidates from O4b: ZTF25aagpypz, ZTF25aafwffi, ZTF24absmrlr, and ZTF24abricne.

To investigate whether the aforementioned transients are nuclear in origin, we employ an improved version of the classifier described in \cite{Pavez.2025}. The original classifier included TDEs as a separate class, while the new version introduces 'Nuclear Transients' (NTs), a broader category encompassing known optical TDEs as well as Ambiguous Nuclear Transients (ANTs). This updated classifier is currently under development (Pávez-Herrera et al., in prep.). Its inputs consist of ZTF alert light curves along with additional metadata from the alerts. This new version also differs from the previous one in that it includes additional features based on multiband light-curve fitting, such as those provided by the Rainbow algorithm \citep{russeil2023rainbowcolorfulapproachmultipassband}. 

\begin{table}[htbp!]
\centering
\renewcommand{\arraystretch}{1.5}
\begin{tabular}{ccc}
ZTF ID & Classification prob & transient type\\
\hline
ZTF23abqkwzr & 0.52 & AGN  \\
ZTF25aagpypz &  0.55& SNII  \\
ZTF24absmrlr & 0.66& NT \\
ZTF24abricne & 0.51& NT\\
ZTF25aafwffi & 0.43& NT\\
\end{tabular}
\caption{Result of nuclear transients classifier for interesting ZTF events with classification probability and the transient type}
\label{tab1}
\end{table}
 The parametric models described in \citealt{Pavez.2025} such as the decay and a machine learning algorithm optimized to identify TDEs \citep[FLEET,][]{Gomez.2023} were updated to follow this multi-band approach by fitting independently to each band, and then the sum of the errors across all bands is optimized while introducing a regularization term that encourages some parameters to remain similar between bands. Furthermore, the probability outputs are more realistic given they have been calibrated using \textsc{CalibrationCV} from \textsc{sklearn}. 

Thus the code classified three of the transients as nuclear transients as represented in table \ref{tab1}. In the case of 'ZTF23abqkwzr' the highest classification probability is AGN with 0.52, the second highest is NT with 0.33. Despite the transient light curve the object was classified as a AGN by ePESSTO+ when the flux was rising \citep{ePESSTO+AGN}.

\subsection{GW-EM Parameter inference}
The GW parameter, specifically the chirp mass, can be estimated using the algorithm described in Section~\ref{mass}. The potential electromagnetic (EM) light curves arising from the motion of the remnant black hole within the accretion disk leading to BHL accretion can be modeled using the framework and equations presented in \citet{Graham_2023}. These models allow us to explore the possible EM signatures associated with compact object mergers embedded in the AGN disks.

The only parameter directly deducible from the LVK public alerts is the chirp mass, which was estimated using the chirp mass prediction   described above. Applying this method to the crossmatched GW events yielded the chirp mass values listed in Table~\ref{tab2}. Details regarding the weights

\begin{figure}[h]
    \centering
    \includegraphics[width=0.45\textwidth]{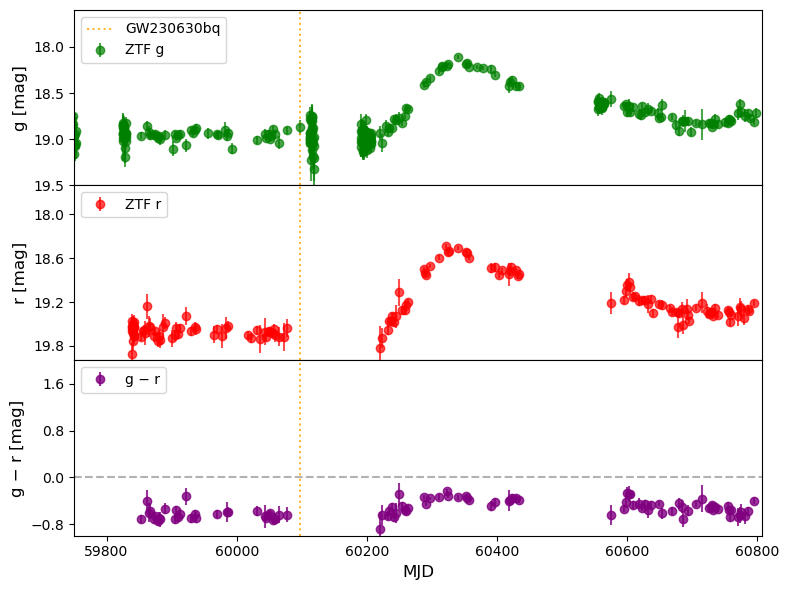}
    \caption{ Light curves of O4a counterpart candidate AT2023zgo/ZTF23abqkwzr in ZTF: g, r bands and g-r color for the  science images and their forced photometry. The vertical line represents the GW event corresponding ZTF alert could be associated with.}
    \label{fig:at2023zgo}
\end{figure}

\begin{figure*}[h]
    \centering
    % --- First row ---
    \begin{subfigure}[b]{0.4\textwidth}
        \includegraphics[width=\textwidth]{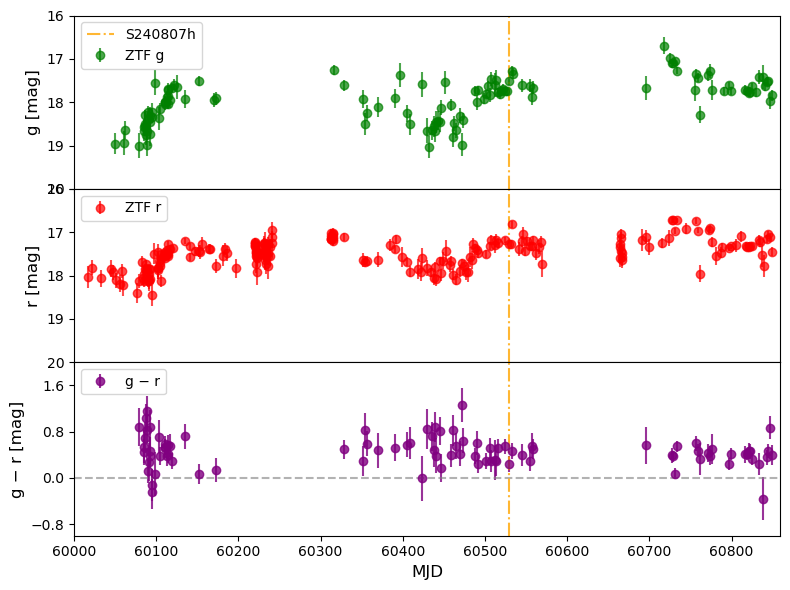}
        \caption{AT2025cze / ZTF25aafwffi}
    \end{subfigure}
    \hfill
    \begin{subfigure}[b]{0.4\textwidth}
        \includegraphics[width=\textwidth]{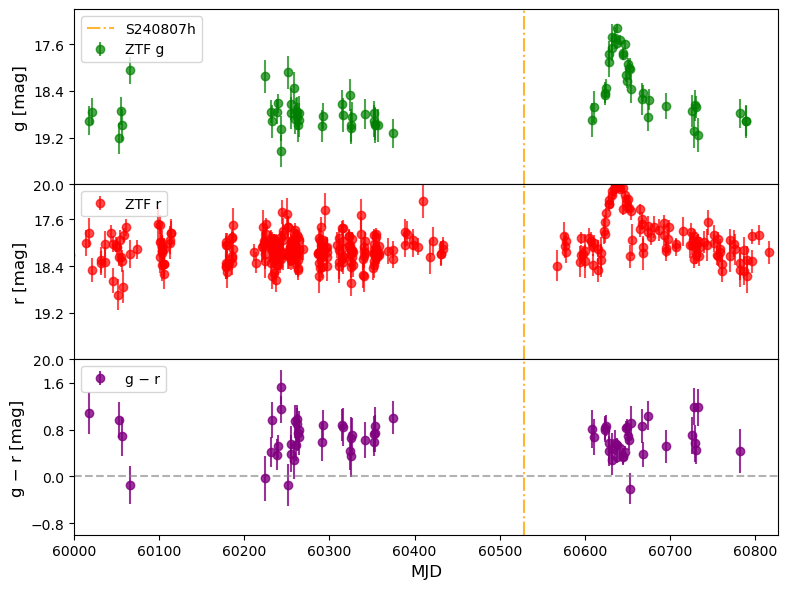}
        \caption{ZTF24absmrlr}
    \end{subfigure}

    % --- Second row ---
    \vskip\baselineskip
    \begin{subfigure}[b]{0.4\textwidth}
        \includegraphics[width=\textwidth]{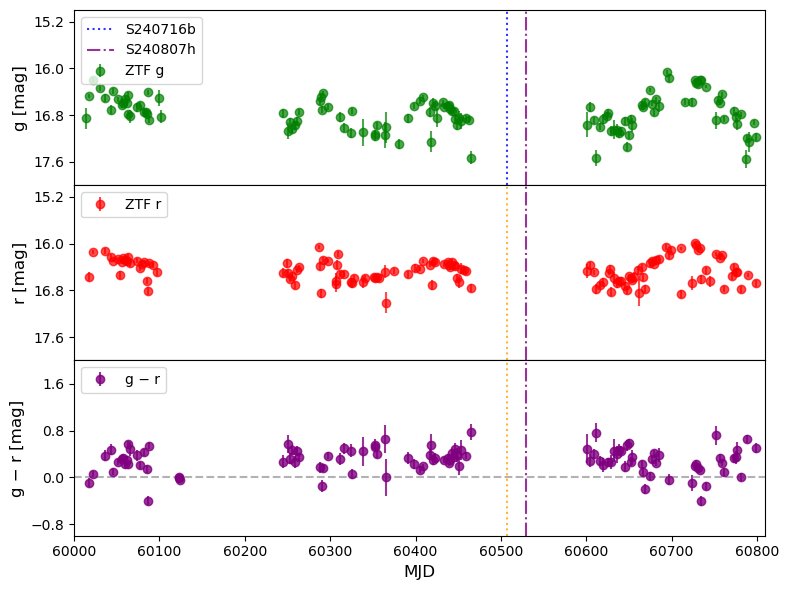}
        \caption{ZTF25aagpypz}
    \end{subfigure}
    \hfill
    \begin{subfigure}[b]{0.4\textwidth}
        \includegraphics[width=\textwidth]{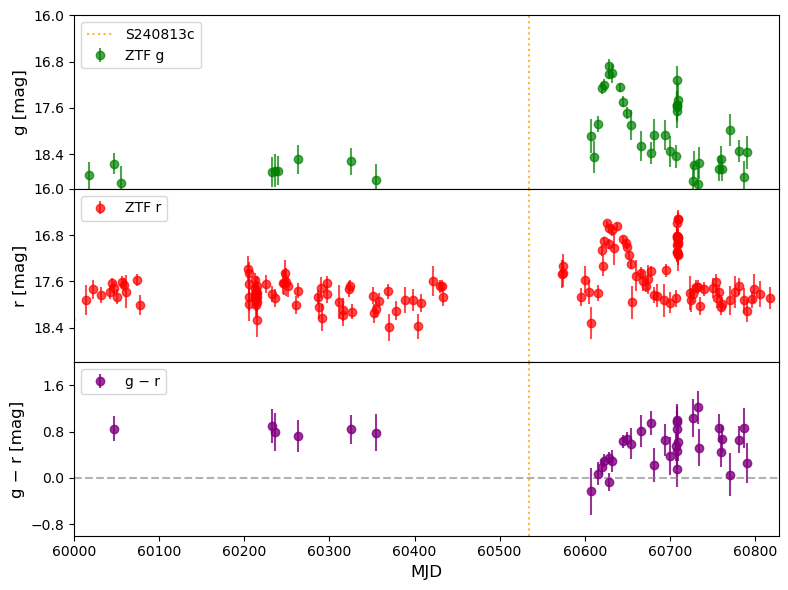}
        \caption{AT2024aaup / ZTF24abricne}
    \end{subfigure}

    \caption{Light curves of O4b counterpart candidates in ZTF: g, r bands and g-r color for the  science images and their forced photometry. The vertical lines in each plot represents the time of the GW event }
    \label{fig:o4bcands}
\end{figure*}

\begin{table}[h!]
\centering
\renewcommand{\arraystretch}{1.5}
\begin{tabular}{ c c }

 Supervent ID & Chirp mass ($M_\odot$) \\
\hline
S230630bq & $9.4^{+5.8}_{-4.3}$ \\
S240807h & $14.6^{+11}_{-6.9}$\\
S240716b & $12.9^{+16.8}_{-4.2}$\\
%S240902bq & $32^{+7.3}_{-7.1}$\\
S240813c &$16^{+9.9}_{-8.9}$\\
\end{tabular}
\caption{GW superevents and their estimated chirp masses using our chirp mass prediction algorithm}
\label{tab2}
\end{table}
The parameter inference for the light curves shown in Figures~\ref{fig:at2023zgo} and~\ref{fig:o4bcands} was performed for the events listed above using the equations presented in \citet{Graham_2023}. The exit timescale, $t_{\mathrm{exit}}$, is defined as the total time between the GW merger and flare peak. This timescale characterizes the period required for the recoiling merger remnant to interact with and emerge from the surrounding disk material, producing observable emission. The rise and decay times of each event were obtained by fitting a model with a Gaussian rise and an exponential decay profile. The maximum recoil velocity was then computed from the total radiated energy of the light curve, assuming a radiative efficiency of 0.1, a disk density of $10^{-10}\,\mathrm{g\,cm^{-3}}$, the SMBH mass, and the observed flare timescale, following the prescription of \citet{Graham_2023}.

Using the maximum recoil velocities estimated for each event, and assuming an optical depth $\tau _{mp}\sim10^{4.5}$ \citep{cabrera.2024} and the inferred exit timescales, we derive disk scale heights of $H/R_{\mathrm{g}} \sim 0.8$--50, corresponding to physical heights of $H \sim 3\times10^{13}$--$1\times10^{14}\,\mathrm{cm}$ for SMBH masses of $10^{7.1}$--$10^{8.4}\,M_{\odot}$. These values represent upper limits on the disk thickness, as they are based on the maximum plausible recoil velocities. The resulting $H/R_{\mathrm{g}}$ ratios are consistent with those expected for the thin to moderately thick regions of AGN accretion disks, as predicted by analytic models \citep[e.g.,][]{Sirko2003,Thompson2005}. The derived parameters for each event are summarized in Table~\ref{tab3}, while a detailed multi-wavelength spectral analysis of the O4a counterpart candidate is presented in Appendix~\ref{a2}. Note that the $g$-band light curves were considered for the parameter inference.

\section{Analysis} \label{analysis}

% \begin{figure}
% \includegraphics[width=8 cm]{ztfpointingso4a.png}
% \centering
% \caption{ZTF pointings for public alerts based on qa product for O4a run duration with $n_{pointings}\geq 15$, where color scale represents the frequency of observations }
% \label{fig:ztfpointo4a}
% \end{figure}
% \begin{figure}
% \includegraphics[width=8 cm]{ztfpointingso4b.png}
% \centering
% \caption{ZTF pointings for public alerts based on qa product for O4b run duration with $n_{pointings}\geq 15$}
% \label{fig:ztfpointo4b}
% \end{figure}

In this section, we evaluate the likelihood of random associations between AGN flares and LVK GW events. Previous studies; such as \citet{Graham_2020}, which reported the first possible association between a GW event and an AGN flare; and \citet{Graham_2023}, which extended this analysis to the full LVK O3 run, have demonstrated the potential for detecting optical counterparts to BBH mergers. In particular, \citet{Graham_2023} identified 20 AGN flares in ZTF Data Release 5 (2018–2021) over a period of $\sim$1000 days, after excluding known supernovae, tidal disruption events, and microlensing events. Among these, seven flares were spatially and temporally coincident with nine GW events from the O3 observing run, which lasted 11 months. Assuming these associations are random coincidences \citep{Veronesi_2024}, we adopt this result as a normalization factor in our study.

\subsection{Simulating random AGN Flares}

\begin{figure}[H]
    \centering
    \begin{subfigure}{0.36\textwidth}
        \includegraphics[width=\linewidth]{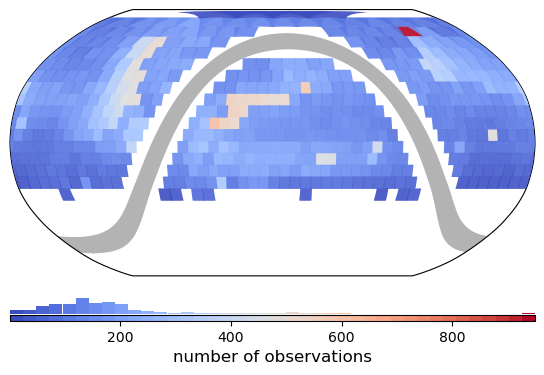}
        \caption{O4a}
        %\label{fig:ztf_a}
    \end{subfigure}
    \hfill
    \begin{subfigure}{0.36\textwidth}
        \includegraphics[width=\linewidth]{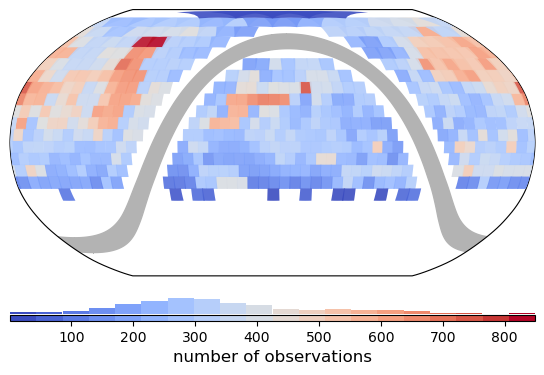}
        \caption{O4b}
        %\label{fig:ztf_b}
    \end{subfigure}
    \caption{ZTF pointings for public alerts based on QA product for O4a (top) and O4b (bottom) with $n_{\rm pointings} \geq 15$, where the color scale represents the frequency of observations.}
    \label{fig:ztfpoints}
\end{figure}

To evaluate the expected rate of chance associations between AGN flares and GW events, we simulated a population of flaring AGNs. We assume that non-GW related AGN flares dominate the population of AGN flares. We randomly select $\sim650$ flaring AGNs from the MQUAS catalog and assign each a flare time (MJD) randomly distributed between the start of the O4 sub-run and up to 200 days after its end. Using the public ZTF survey pointing history (Figure~\ref{fig:ztfpoints}), we assessed, for each LVK skymap from the O4 sub-runs, whether a given flare would have been detected within the observed fields. This is within a 200-day time frame post-merger, and $2\sigma$ distance limit of the respective GW event inferred distance. 

\subsection{Expected Number of flares}
 We start by measuring the number of randomly associated flares as a function of the GW skymap area. To do this, we measure the number of detected flares among those simulated for different GW skymap area thresholds. These values are normalized relative to those reported by \citet{Graham_2023}, i.e., so that we recover about 20 flares in AGNs of MQUAS per 1000 days independent of the GW events.

 The simulated detection statistics were then compared with the observed events (mentioned in the previous section) in terms of the cumulative number of detections as a function of the GW skymap area threshold (see Figure~\ref{fig:sims}). The comparison reveals an excess in the number of detected events relative to the random expectation, with one excess event identified during the O4a run and four excess events during the O4b run. In O4a, if we consider a GW skymap area threshold of 2,000 deg$^2$, the probability of having one or more random associations is $7.6\pm 2.5$\% (1$\sigma$). In O4b, if we consider a GW skymap area threshold of 15,000 deg$^2$, the probability of having three or more random associations is $0.6 \pm 0.3\%$ (1$\sigma$). This suggests that we can marginally discard the possibility that the observed events are random associations in O4b superevents.  

 \subsubsection{O4b skymap area distribution}
 
We further statistically test and analyze the O4b real and simulated event distributions, to strengthen or discard our hypothesis. Instead of only comparing the expected number of events for a given threshold skymap area, we test whether the distribution of AGN flare associated skymap areas resemble those from O4b.

For this, we introduced a slight modification to the simulation framework: AGN flares were uniquely associated with GW events through a likelihood-based matching scheme. This ensures a physically consistent, one-to-one association between AGN flares and GW events, while incorporating both spatial and distance information into the likelihood score. This allowed a more realistic evaluation of the recovered event distributions and reduced contamination from false or ambiguous matches.

The likelihood score quantified the probability of a physical association between an AGN and one or more GW events, defined as the product of (i) the AGN’s sky-location probability density within the GW skymap and (ii) the consistency (Gaussian probability weight) between its redshift-inferred distance and the GW event’s luminosity-distance posterior. AGNs with the highest scores were assigned exclusively to their most probable GW counterpart, while weaker associations were discarded. 

The empirical cumulative distribution functions (CDFs) for the skymap areas in the O4b run, for simulated and for observed events are shown in Figure 
\ref{fig:kstest}.
Although there seems to be an excess from large skymap areas, indicative of chance coincidences, a two-sample Kolmogorov–Smirnov (KS) test comparing the observed and simulated skymap area distributions indicates that they are statistically consistent with being drawn from the same parent population (p = 0.45). Thus, we cannot discard the hypothesis that the observed events follow the distribution of simulated events. However, note that this comparison is valid for the extreme case where all flaring events are associated with GW events. A more realistic approach would be to visually inspect all the AGN flares, but this cannot be done in our simulation.

In addition to the Poisson and KS tests, we performed a Mood’s median test to compare the central tendency of the detection area distributions between simulations and observations. The test yielded a p-value of 0.26, indicating no statistically significant difference in the median detection areas. 

\subsection{Recovering BH merger triggered flares}

We now consider the case where all GW events trigger an AGN flare and check whether their optical counterparts would be detected. For this, we can use the GW event distance and chirp mass (available only for O4a) to estimate the expected apparent magnitude of the flare.

Using the GWTC-O4 catalog \citep{lvk.2025}, we obtained the chirp masses of the reported binary mergers in O4a and estimated their corresponding kick velocities arising due to mass asymmetry (assuming no spin) from equation from 4 in \citealt{Baker.2008}, approximated as below:
\begin{equation}
    v_{kick} \approx A\eta^2(\sqrt{1-4\eta})(1+B\eta),
\end{equation}
where $v_{\mathrm{kick}}$ is the kick velocity (in $\mathrm{km} \mathrm{s^{-1}}$), $\eta$ is the symmetric mass ratio $\frac{m_1m_2}{(m_1+m_2)^2}$, $A=1.35\times10^4\ \mathrm{km}\ \mathrm{s^{-1}}$ and $B=-1.74$. The optical luminosity of the kicked remnant was then computed using Equation (3) in \cite{Graham_2023}, which was subsequently converted into AB magnitudes.

 A similar simulation was performed following the same procedure described previously, with the modification that the flaring AGN distance distribution was now sampled from the luminosity-distance distributions of the GW events.  Each sampled event was assigned a realistic apparent magnitude, m$_{\mathrm{AB}}$, and the detection threshold was imposed based on the limiting magnitude of ZTF. When assigning an apparent magnitude we assume an AGN disk density of $10^{-10} \mathrm{g}\ \mathrm{cm^{-3}}$ as in \cite{Graham_2023}.  Under the previous assumption, all of the GW triggered flares would have been detected in O4a. 

\begin{figure}[H]
    \centering
    \begin{subfigure}{0.34\textwidth}
        \includegraphics[width=\linewidth]{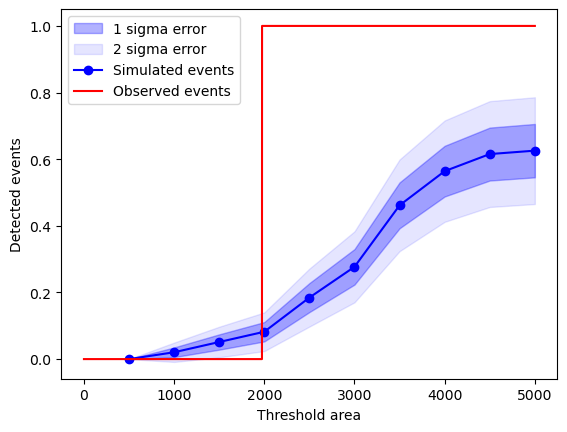}
        \caption{O4a}
        \label{fig:sim_o4a}
    \end{subfigure}
    \hfill
    \begin{subfigure}{0.34\textwidth}
        \includegraphics[width=\linewidth]{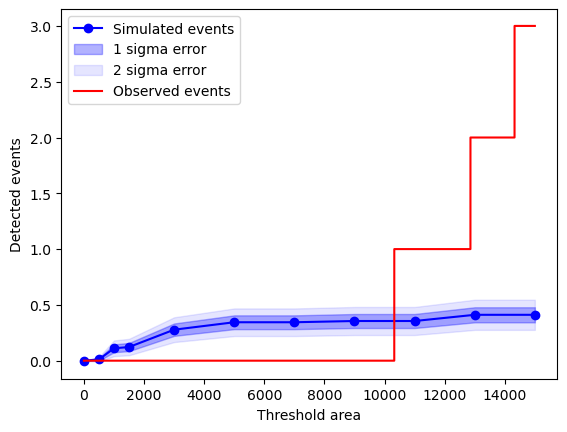}
        \caption{O4b}
        \label{fig:simo_b}
    \end{subfigure}
    \caption{Cumulative distribution function of the simulated (in blue) and detected flare events (in red) corresponding to ZTF pointings up to 200 days post merger for each GW event as a function of GW skymap area. The top figure corresponds to O4a, and the bottom, O4b.}
    \label{fig:sims}
\end{figure}

To summarize, we initially evaluated the likelihood of random associations using a Poisson test. This indicated that the single event detected during the O4a run could plausibly be a random association, whereas the four events detected during O4b might correspond to real physical connections.

\begin{figure}[H]
\includegraphics[width=8cm]{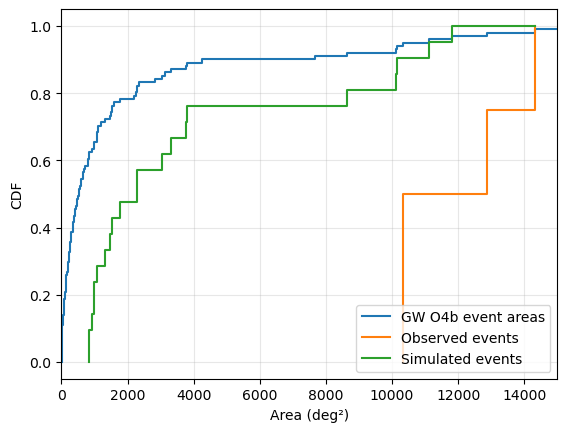}
\centering
\caption{Cumulative distribution functions (CDFs) of 90\% localization areas for the O4b GW events (blue), simulated events (green), and observed AGN flare associations (orange). The median localization areas are 545 deg² for the O4b events, 2,300 deg² for the simulated sample, and 11,590 deg² for the observed sample.}
\label{fig:kstest}
\end{figure}

\begin{table*}[h!]
\centering
\renewcommand{\arraystretch}{1.3}
\resizebox{\textwidth}{!}{%
\begin{tabular}{l c c c c c c c c c c c c}

ZTF ID & \textbf{RA} & \textbf{Dec} & Superevent ID & prob & redshift & $t_{\mathrm{exit}}$ & $t_{\mathrm{rise}}$ & $t_{\mathrm{fall}}$ & $\log_{10}E$ & $\log_{10}(M_{BH})$ & $v_{k,\max}$ & $H_{\max}$ \\
 & (h:m:s) & (d:m:s) &  &  &  & (days) & (days) & (days) & (erg) & ($M_\odot$) & (km\,s$^{-1}$) & [$r_g$] \\
\hline
ZTF23abqkwzr & \textbf{08:48:54.89} & \textbf{+31:47:00.9} & S230630bq & 0.41 & 0.232 & 133 & 61 & 497 & 51.67 & 7.41 & 156 & 10.4 \\
ZTF25aafwffi & \textbf{16:08:31.61} & \textbf{+07:08:18.2} & S240807h & 0.76 & 0.153 & 188 & 16 & 87 & 50.06 & 7.13 & 304 & 54.2 \\
ZTF25aagpypz & \textbf{11:09:19.79} & \textbf{+71:42:33.8} & S240716b & 0.83 & 0.130 & 191 & 84 & 59 & 50.71 & 8$^*$ & 213 & 5.2 \\
ZTF25aagpypz & \textbf{11:09:19.79} & \textbf{+71:42:33.8} & S240807h & 0.79 & 0.130 & 169 & 84 & 59 & 50.71 & 8$^*$ & 231 & 5.0 \\
ZTF24absmrlr & \textbf{08:01:15.68} & \textbf{+53:18:58.7} & S240807h & 0.88 & 0.126 & 106 & 12 & 39 & 49.98 & 8.05 & 203 & 2.5 \\
ZTF24abricne & \textbf{08:58:04.62} & \textbf{+12:28:21.6} & S240813c & 0.80 & 0.117 & 92 & 14 & 25 & 50.14 & 8.38 & 174 & 0.86 \\

\end{tabular}%
}
\caption{ZTF events and corresponding GW superevents with location and light-curve properties. SMBH masses are estimated using \texttt{ppxf} \citep{Cappellari2023} fits to SDSS spectra and the virial mass relation of \citet{Greene_2005}. A fiducial $\log_{10}(M_{BH}) = 8$ is adopted when no archival spectra are available (indicated by *).}
\label{tab3}
\end{table*}

 To further investigate the O4b case, in particular their large skymap areas, we applied KS and Moore median tests. The simulated and observed events are consistent with being drawn from the same underlying distribution. This implies that the skymap areas seen in the distribution of observed events are not significantly larger than those from the simulations.
 
 Additionally, if AGN accretion disk gas densities are $< 10^{-10} \mathrm{g} \mathrm{cm^{-3}}$ and recoil velocities dominated by mass asymmetry, our simulations indicate that all such events, if  occurring within an AGN disk, would be detectable.

\section{Discussion} \label{discussion}

We developed an algorithm using the ALeRCE infrastructure that, upon receiving GW skymaps, applies a sequence of spatial, temporal, and probabilistic filters to identify promising electromagnetic counterpart candidates. We applied this algorithm to multi-detector GW superevents from the O4a and O4b observing runs, identifying one and four interesting candidates, respectively. Several of these have already been reported to the TNS, though they remain unclassified. Two of the four events lie in the overlapping regime of SN and TDE (ZTF24abricne, ZTF24absmrlr); one in the TDE regime (ZTF25aafwffi), and two in the AGN regime (ZTF23abqkwzr, ZTF25aagpypz) as defined in Figure 2 of \citet{Graham_2023}.

To place the inferred AGN disk parameters (Table~\ref{TAB2}) in the context of AGN disk theory, we note that the derived scale heights, $H/R_{\mathrm{g}} \sim 0.8$--50 for SMBH masses of $10^{7.1}$--$10^{8.4}\,M_{\odot}$, were obtained using the maximum recoil velocities and therefore represent upper limits on the disk thickness. Most values ($H/R_{\mathrm{g}} \lesssim 10$) are consistent with the thin or moderately thick regions of AGN accretion disks predicted by analytic models \citep[e.g.,][]{Sirko2003,Thompson2005}, which yield $H/R \sim 10^{-2}$--0.3 for the inner to self-gravitating disk regimes. The few larger ratios likely reflect either the assumption of maximum kick velocities or local disk inflation. Overall, the inferred geometries, even as upper limits, are broadly consistent with the vertical structures expected for AGN disks, supporting an origin of these flares within the inner or moderately thick disk regions. If the recoil velocity can be independently constrained from GW data \citep{vijay.2019,vijay.20}, this approach could be used to estimate the disk density of the corresponding AGN accretion disk.

We also estimated the expected number of optical counterparts arising from random flare associations within ZTF alerts during the O4a and O4b runs. This estimate, shown as a function of GW skymap threshold area in Figure~\ref{fig:sims}, provides a baseline for interpreting the statistical significance of our detections. The observed excess, in context of the observed event areas being larger than the median GW skymap areas (as shown in \ref{fig:medarea}) suggests a random association in O4a, whereas the four events detected during O4b might correspond to real physical connections.

As suggested by \citet{Graham_2023} and possibly demonstrated in \citet{cabrera.2024}, genuine counterparts are expected to exhibit spectra consistent with an off-center explosive event, potentially producing asymmetric broad-line components. However, the O4a candidate shows no such asymmetry in its broad-line profile, and for the O4b candidates we were unable to test this spectroscopically, as the flare timescales were shorter and the events were identified well after their peak activity. While some of these sources could represent false positives, the use of the \textit{nuclear transient classifier} helps to mitigate this risk by selecting variability associated with AGN.
%We also excluded any events already classified as TDEs, SNe, or other known transients in TNS.

\subsection{Implications for Upcoming Observing Runs}

Our algorithm was designed within the ALeRCE alert broker framework but can be readily adapted for use with other brokers, if similar classification probabilities are provided. Running it at least once per month during future observing runs would allow detection of transients during their rising phase. Leveraging ALeRCE's \textit{watchlist} \footnote{\url{https://alerce.science/services/watchlist/}} feature enables continuous tracking of high-priority GW counterpart candidates, facilitating timely spectroscopic follow-up. This approach provides an efficient, survey-based means to identify potential electromagnetic counterparts to GW events with large localization areas particularly valuable when dedicated follow-up facilities are limited. Thus, our method offers a complementary route to capture transient candidates that might otherwise be missed.

\subsection{Comparison with Other Studies}

\citet{cabrera.2024} reported a probable electromagnetic counterpart to the GW event S230922g based on dedicated follow-up observations. Since this source lies too far south to be observed by ZTF, its location likely explains why our search did not recover the candidate. \citet{sun2025at2021aeukrepeatingpartialtidal} discussed the possibility of a repeated partial TDE, i.e., AT2021aeuk being a result of a stellar mass merger within the accretion disk that could produce similar phenomena.

\citet{darc2025longtermopticalfollows231206cc} outlined a framework for long-term optical monitoring of BBH events, while \citet{zhang2025jointsearchelectromagneticcounterpart} emphasized coordination between wide-field imagers and multi-object spectrographs to maximize multi-messenger yield.

More recently, \citet{cabrera2025multimessengerconstraintsligovirgokagragravitational} statistically expanded upon the results of \citet{Graham_2023}, demonstrating that fewer than 3\% of BBH mergers are likely to produce an observable optical counterpart. This implies that up to $\sim$40\% of BBH mergers could occur within AGN accretion disks, though only a small fraction would yield detectable flares. These findings are broadly consistent with our results, where approximately 1 out of 70 events in the O4a run and 3 out of 105 events in the O4b run exhibited potential optical counterparts. 

Furthermore, \citet{cabrera2025multimessengerconstraintsligovirgokagragravitational} found no compelling evidence that higher-mass mergers are preferentially associated with AGN flares, suggesting that the apparent excess of high-mass events reported by \citet{Graham_2023} may instead arise from chance coincidences. Our analysis explores around the scenario that the AGN flares could be linked to lower chirp mass and distant mergers, a scenario that remains compatible with the inferred occurrence rates.

\section{Conclusions} \label{conclusions}
Establishing a robust association between LVK GW events and possible optical flare counterparts remains a significant challenge in the pursuit of identifying BBH-EM counterparts. Given that such transients occupy a region of parameter space overlapping with TDEs and other nuclear transients, the development of improved theoretical models and time-dependent simulations for slowly evolving transients is essential. Dedicated follow-up campaigns of LVK events, such as those presented in \citet{cabrera.2024}, will be critical for advancing this effort.

We have developed an algorithm that leverages the ALeRCE infrastructure to systematically search for optical counterparts to GW superevents detected during the LVK O4a and O4b observing runs. The search yielded five promising candidates, one of which was previously reported as AT2023zgo on the Transient Name Server (TNS), while the remaining sources are yet to be conclusively classified.

We find a modest excess of AGN flares spatially and temporally coincident with gravitational-wave localizations compared to random expectations, particularly during the O4b observing run. Although these coincidences occur predominantly in superevents with large localization areas where random overlaps are most likely. Our algorithm is specifically designed to efficiently identify potential associations over such extended regions, providing sensitivity to counterparts that may otherwise remain undetected. Statistical comparisons between the observed and simulated samples using KS and Mood’s median tests yield no significant differences, indicating that the observed events are statistically consistent with expectations for genuine AGN–GW associations. Overall, the evidence remains suggestive rather than definitive, underscoring the need for larger samples, improved localization, and detailed multi-wavelength follow-up to confirm or refute a physical connection between AGN flares and GW events.

Nonetheless, our method remains highly effective in probing large sky areas that are typically inaccessible to targeted telescope follow-up. By statistically prioritizing AGN candidates based on probabilistic consistency rather than a brute force approach, this approach provides a complementary path to GW EM counterpart searches.

Furthermore, the imminent dissemination of public alert streams from LS4 \citep{Adam.2025} and the Vera C. Rubin Observatory’s \citep[LSST,][]{Ivezic.2018} will dramatically expand the discovery space for optical transients. The resulting increase in event volume will significantly enhance the prospects for identifying optical counterparts to BBH mergers and for performing systematic, population-level studies of nuclear transients potentially associated with GW sources. In this context, our framework provides a baseline and statistically driven foundation for prioritizing multimessenger candidates in real time, ensuring readiness for the high-cadence, data-rich environment of forthcoming survey facilities, however further scalability tests need to performed.

\begin{acknowledgements}

Thanks to Amrita Singh for comments and help reading the manuscript. 
H.B. acknowledges the financial support from ANID NATIONAL SCHOLARSHIPS/DOCTORATE 21241862.
F.F. and H.B. acknowledge support from ANID BASAL Project CATA (FB210003).
A.M.A., F.F. and H.B. acknowledge support from ANID BASAL Project CMM (FB210005).
A.M.A., F.F., H.B. and M.L.M.A. acknowledge support from ANID Millennium Science Initiative MAS (AIM23-0001). L.H.G. acknowledges financial support from ANID program FONDECYT Iniciaci\'on 11241477.
M.L.M.A. and L.H.G. acknowledge support from ANID Millennium Science Initiative TITANS (NCN202\_002).
M.L.M.A. acknowledges support  from the China-Chile Joint Research Fund (CCJRF2310).
      
\end{acknowledgements}
\section*{Data Availability}
All source codes and trained models used in this work (apart from nuclear transients classifier, which will be available in a future publication) are publicly available.
The ALeRCE stamp and light-curve classifiers were not implemented in this work, we obtained the corresponding classification probabilities from the ALeRCE database using the direct SQL access. The ALeRCE pipeline repository is found in GitHub \footnote{\url{https://github.com/alercebroker/pipeline}}
.
The GW–AGN Watcher framework, including the cross-matching pipeline and chirp-mass prediction algorithm, is available in GitHub \footnote{\url{https://github.com/Hemanthb1/GW_AGN_watcher}}, archived at Zenodo \footnote{\url{https://doi.org/10.5281/zenodo.18742242}}and also as a pip installable package\footnote{\url{https://pypi.org/project/gw-agn-watcher/}}.
This repository includes documentation, example notebook, AGN catalog and configuration files required to reproduce the results presented in this paper.
\section*{Software and Facilities}
This research made use of \texttt{Astropy} \citep{astropy:2013, 
astropy:2018, astropy.2022}, \texttt{HEALPix} \citep{Gorksi.2005}, \texttt{ligo.skymap} \citep{Singer_2016}, \texttt{Matplotlib} \citep{matplotlib}, \texttt{NumPy} \citep{numpy}, \texttt{Pandas} \citep{pandas}, PostgreSQL \citep{postgresql_global_development_group_postgresql_2022}, \texttt{SciPy} \citep{scipy}, \texttt{Scikit-learn} \citep{scikit-learn}, \texttt{alphashape} \citep{bellockk_alphashape}, \texttt{ALeRCE} \citep{forster2021}, \texttt{gracedb-sdk} \citep{singer_gracedb-sdk_2022}, and \texttt{pesummary} \citep{pesummary2020}, Q3C \citep{2006ASPC..351..735K}. We also used data and software from the \textit{Zwicky Transient Facility (ZTF)} \citep{Bellm_2018}, the \textit{LIGO–Virgo–KAGRA Collaboration (LVK)} \citep{abbott.2016} observing runs and sky localizations.

\bibliographystyle{aa}
\bibliography{bibliomain}

\begin{appendix}

 \section{chirp mass prediction}
This section contains plots representing testing and validation results of our chirp mass prediction algorithm for O3 merger events
    \begin{figure}[h!]
    \includegraphics[width=8cm]{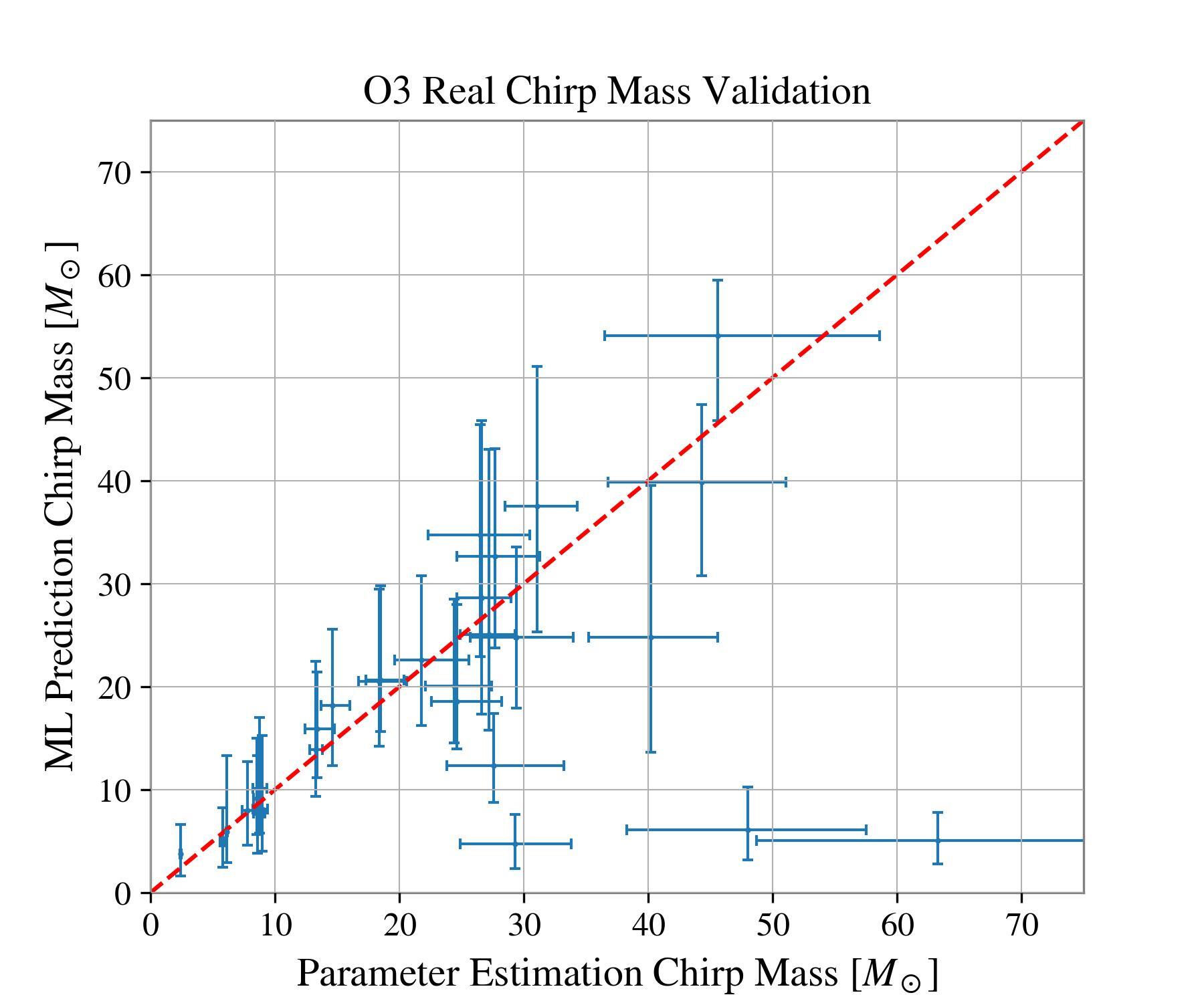}
    \centering
    \caption{Performance of the algorithm on the GWTC-2.1 \citep{abbott.2024} event catalog. The X axis denotes chirp mass inference for each event from LALInference \citep{Veitch_2015} parameter estimation and the Y axis represents masses predicted by our OCP algorithm. These estimations were made using an identical number of BAYESTAR simulations made with the actual O3 detector noise curves. Some events are very poorly estimated in the low mass region; this can be attributed to variability in the stability of detector noise curves.}
    \label{fig:algper}
    \end{figure}

    \begin{figure}[h!]
    \includegraphics[width=8cm]{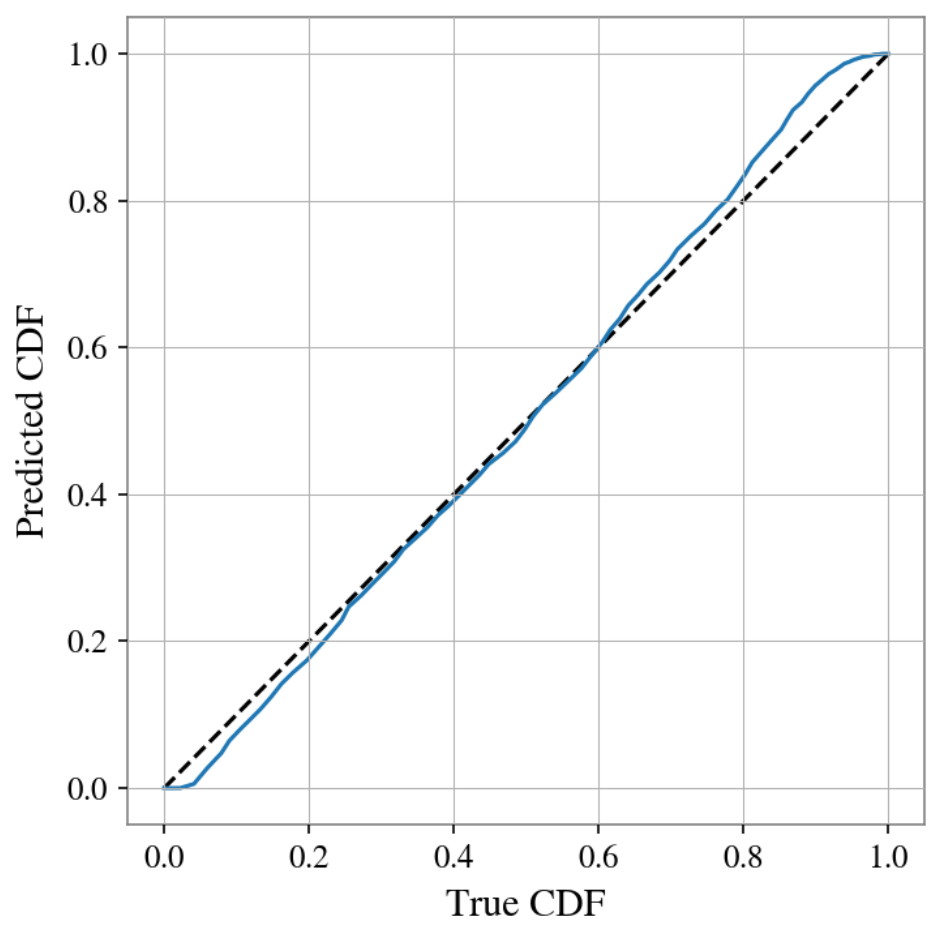}
    \centering
    \caption{P-P plot of predicted chirp mass vs true chirp mass of simulated events. The black dashed line represents ideal situation with no bias, while blue line represents the pp curve of the algorithm. The distributions are similar to each other, with small bias occurring at the edges of the distribution.}
    \label{fig:algcdf}
    \end{figure}
    \newpage

 \section{Multiwavelength follow up of AT 2023zgo}\label{a2}

\begin{figure}[ht!]
    \includegraphics[width=8cm]{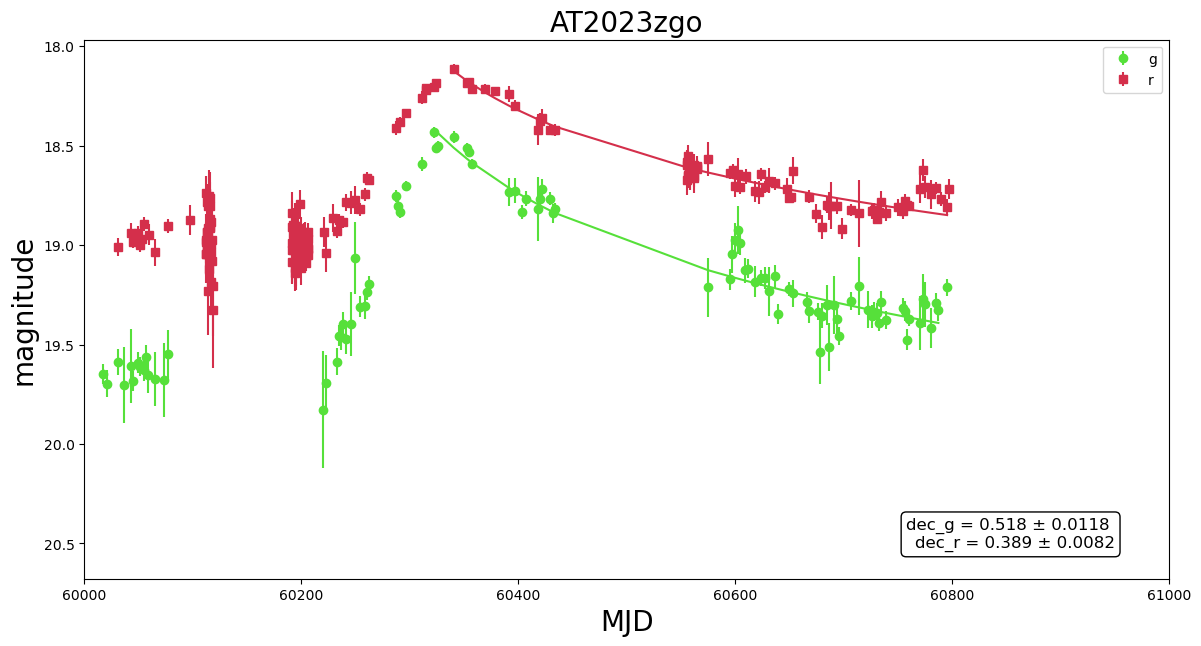}
    \centering
    \caption{Fitting the light curve with a power-law decay model(TDE) with reference to \citep{Pavez.2025}. The a power law index fit values are of 0.518 $\pm 0.0118$ for g-band and 0.389 $\pm$ 0.0082 for r-band in contrast to -1.66 expected for a TDE. }
    \label{fig:tdefit}
    \end{figure}

 In this section, we present the multiwavelength follow-up and analysis of the event AT2023zgo. The optical light curve was tested for consistency with a TDE using a power-law decay fit. However, as shown in Fig.~\ref{fig:tdefit}, the best-fitting power-law model does not reproduce the observed evolution, indicating that AT~2023zgo is inconsistent with a TDE interpretation.

    Given the uniqueness of the flare and its potential association with a binary black hole (BBH) merger, we examined the source for emission in the ultraviolet (UV) and X-ray regimes to search for possible jet-related signatures, as relativistic jets have been proposed in BBH merger environments \citep{Tagawa_2023}. 
  The follow-up observations for other potential counterpart candidates were not feasible due to the short timescale and rapid decline of those transient events. Future coordinated multiwavelength campaigns will be essential to confirm the physical nature of Bowen-type flares and to distinguish them from other classes of AGN-related transients.
 %The O4a counterpart candidate AT2023zgo was discovered by our algorithm while it was still in development. The timescale of the flare of AT2023zgo was so large, and we identified the candidate during its decaying phase (around 330 days post peak flare)
 \subsection{Bowen Fluorescence lines}
    
    \cite{Trakhtenbrot2019} described the discovery of the transient optical event AT 2017bgt, which was observed in the nucleus of the early-type galaxy 2MASX J16110570+0234002, at $z$=0.064. This event was characterized by an unusual optical flare and the appearance of Bowen fluorescence flares (BFF) in its spectrum. The authors suggested that the SMBH experienced a sudden increase in their UV-optical emission.
    
    When high-energy UV radiation from a central source, like a supermassive black hole in an AGN, ionizes atoms or ions in the surrounding gas, this process excites the gas, causing it to emit light as it returns to lower energy states. Specifically, the 
    photons ionize helium (He II $\lambda$304$\AA$),  which then induces a cascade of electronic transitions in ions such as O III (with emission lines around $\lambda$3133 Å) and N III (with emission lines around $\lambda$4634 Å, $\lambda$4641 Å), producing distinctive emission lines known as Bowen fluorescence \citep{bowen1928}.
    
    \cite{makrygianni2023} presented a multiwavelength analysis of AT 2021loi, which showed the first clear evidence of a rebrightening after around one year after the first flare in a BFF. This source shows an optical rise of a factor four in about one month, and a slow decline of 13 months, and then the rebrightening after 400 days. They discussed that the event could be related to an increase in the Eddington rate.
    
    AT 2019aalc is another BFF that showed a first peak in 2019, and a rebrightening in 2023 \citep{milanveres2024, sniegowska2025}. The rebrightening indeed showed stronger flux increase than the first peak, strong and broad ($\sim$2000 km s$^{-1}$) He II $\lambda$4686$\AA$ and N III $\lambda$4640$\AA$ lines, a slow decay of the light curve in general agreement with other BFFs. The first peak was coincident with a neutrino event \citep[IC\,191119A][]{icecube2019}. \cite{ sniegowska2025} discussed that the origin of the BFF could be related to radiation pressure instabilities in the (pre-existing) accretion disk of an AGN.

    Other sources that have been related to the BFF family are F01007 \citep{tadhunter2017}, 
    OGLE17aaj \citep{gromadzki2019},
    and AT2019avd \citep{malyali2021}.

    One curious property of almost all these systems is that they occurred in Narrow Line Seyfert 1 (NLSy1) galaxies.

    BFFs have also been observed in TDEs \citep[e.g.,][]{leloudas2019}, but their properties seem to differ from these BFF systems, which show slow decline in their light curves, a rebrightening after years, the absence of very broad He II $\lambda$4686$\AA$ ($>$10000 km s$^{-1}$), and have preference to occur in AGN.
    
 \subsection{Spectroscopic observations}

\begin{figure}[h!]
\includegraphics[width=8cm]{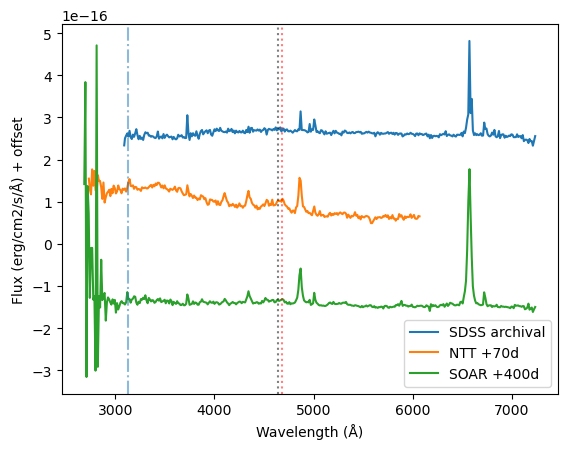}
\centering
\caption{AT2023zgo spectroscopic follow-up observations - Bowen Fluorescence lines OIII(3133 Å),HeII(4696 Å), NIII(4640 Å) overplotted against candidate of GW event S230630bq (+130d pm)}
\label{fig:spec}
\end{figure}

Light curves and spectra play a fundamental role in distinguishing among the diverse transient events observed in AGNs, including tidal disruption events and changing-look events. In light of the peculiar characteristics of the transient AT2023zgo, as reported in the TNS, we conducted a Target-of-Opportunity (ToO) spectroscopic follow-up using the SOAR Goodman spectrograph (Program ID: SOAR2025A-027, PI: H. Bommireddy). To obtain full coverage of the optical wavelength range, we employed the Goodman Red configuration in both the 400 M1 and 400 M2 modes, with a total exposure time of 1200 s per configuration. Arc lamp and spectrophotometric standard star calibrations were acquired with the science observations. The two-dimensional Goodman spectra were reduced using standard steps, including bias removal, flat-field correction, and wavelength calibration using NeArHg lamps. The wavelength calibration was checked against sky emission lines. The one-dimensional spectra were optimally extracted and then flux-calibrated using a flux standard observed immediately after the science spectra  \citep[see][]{Cartier_2026}, for more details). The resulting spectra is shown in green in Figure \ref{fig:spec}.

The archival spectra (in blue) of the host galaxy was retrieved from SDSS DR17 (MJD: 52991), the reported spectrum in TNS as on peak brightness (~70 days after detection - MJD 60318) of the event with NTT \citep{ePESSTO+AGN}, and with SOAR ($\sim 400$ days after detection -MJD 60699)

    \subsection{Fitting of optical spectra}
    Spectra were corrected for Galactic extinction following the law of \citet{cardelli89} and later corrected by redshift ($z=0.23296$). In the spectral fitting model includes contributions from the accretion disk (modeled as a power-law), the stellar population, iron pseudo-continuum \citep{boroson92} and all emission lines reported by \citet{vander.2001} plus the Bowen Fluorescence line NIII$\lambda4640$. Among the three available spectra, the SDSS observation corresponds to the dimmer AGN activity. We therefore used this spectrum to model the stellar contribution, employing the pPXF code \citep{cappellari2017} and the E-MILES stellar population (SSP) templates \citep{vazdekis2016}. Given that the stellar component evolves on timescales higher than $10^6$ years, it was fixed in the spectral fitting of the NTT and SOAR observations, allowing only the powerlaw component to vary. All the emission lines were modeled with Gaussian profiles, and a double Gaussian profile was considered to reproduce both broad and narrow components, as in the case of the Balmer lines. The final spectral fitting was obtained using a MCMC-based fitting code employed in previous AGN studies. 
    
\begin{figure}[h!]
    \includegraphics[width=8.5cm]{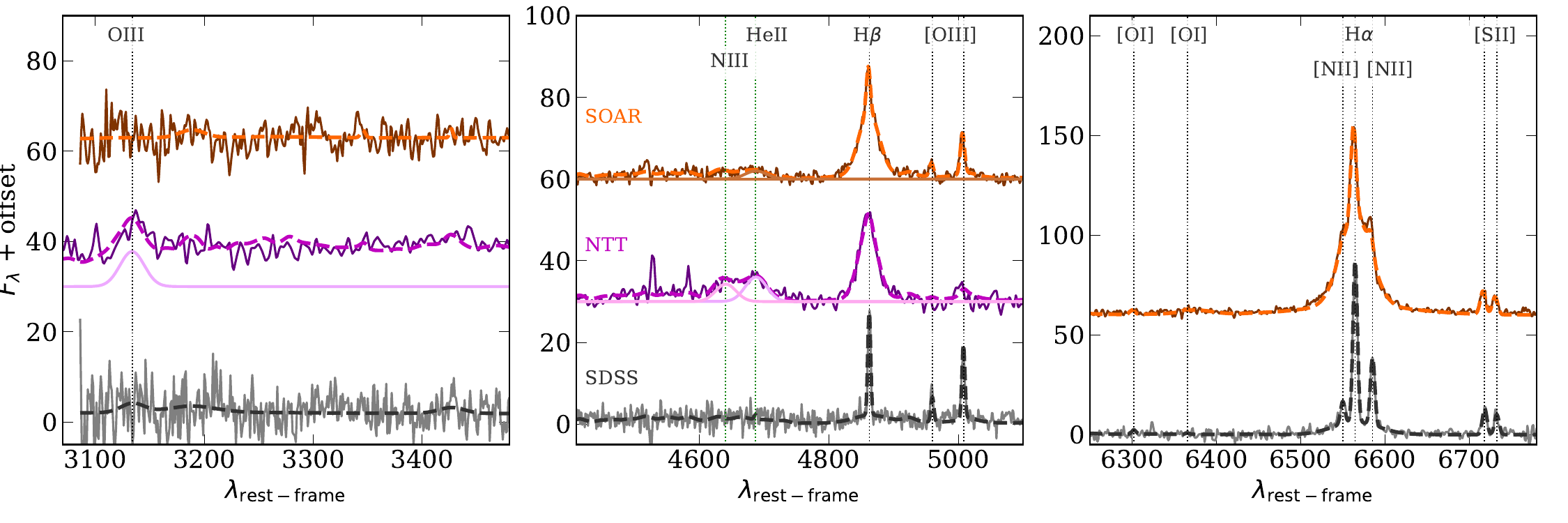}
    \centering
    \caption{Comparison of the relevant spectral regions after continuum subtraction for each observation. A constant factor has been added to the NTT (+30) and SOAR (+60) spectra for visualization purposes. Dashed line correspond to the total fitting. The vertical lines indicates the wavelength rest-frame of the emission lines.}
    \label{fig:zoom}
    \end{figure}

    The equivalent widths ($W$) and the Full Width at Half Maximum (FWHM) of the most important emission lines are reported in Tab.~\ref{table:measurements}, while a comparison of the relevant spectral regions (after continuum subtraction) is shown in Fig.~\ref{fig:zoom}.
    %The spectra fitting decomposition is shown in  Fig.~\ref{fig:xxxx}. 
    We found that the power-law becomes steeper in the NTT and SOAR spectra, indicating a significant increment in the UV ionizing continuum flux and the appearance of the Balmer bump in the spectrum after the flare. This coincides with the remarkable increment in the equivalent widths of the Balmer lines for the NTT and SOAR observations compared to the dimmer state (SDSS spectra). The detection of the Bowen Fluroscence emission lines such as OIII$\lambda$3133 and NIII$\lambda4640$, as well as high ionization lines like HeII$\lambda4685$ and HeI$3188$,  also support the presence of a harder UV continuum in the post-flare spectra. The FWHM measured for OIII$\lambda$3133 and NIII$\lambda4640$ is similar to the one found in the Balmer Lines (2000-3000 \kms). This suggests that the Bowen Fluorescence emission lines originate also in the broad line region (BLR). 
    
    To model the H$\alpha$ emission line in the SOAR spectrum, we included a second broad Gaussian profile to reproduce the base of the emission line. This component shows an equivalent width of 75.1$\pm$4.1 {\rm \AA} and a FWHM of 6,954$\pm$75 \kms. Considering both components, the FWHM of the total profile is 2049$\pm$112 \kms. We also include a second broad component in the H$\beta$ profile, however, its contribution is negligible. The physical meaning of a second component in the H$\alpha$ profile would indicate an increment in the velocity dispersion of the BLR gas or the presence of outflows. We measure the velocity centroid\footnote{$c_x= \frac{\lambda_{R(x)}+\lambda_{B(x)}-2\lambda_0}{2\lambda_0}c$,
    where  $\lambda_0$ is the rest-frame wavelength, $\lambda_{R(x)}$ and $\lambda_{B(x)}$ are the red and blue wavelengths at $x=\frac{1}{4}, \frac{1}{2}, \frac{3}{4}$, $\frac{9}{10}$, and $c$ is the speed of light.} ($c_x$) at different level of the profile investigate possible asymmetries, a key indicator of a BBH merger. The velocity centroid is $\sim30$ \kms at different levels of the profile, corresponding to a shift of $<1\AA$ with respect to the H$\alpha$ rest-frame. This indicates that the full H$\alpha$ profile is completely symmetric and rule out the presence of any outflow. So, we conclude that the third broad component corresponds to an increment in the velocity dispersion of the BLR. 
    
    Table~\ref{table:measurements} lists the bolometric luminosity, black hole mass and Eddington ratio. The luminosity of the AGN at 5100$\AA$ is estimated from the powerlaw fitting. %The luminosity in the thee observations is simila, why?? ¿los espectros están normalizados? el nivel del continuo deberia cambiar y no se ve eso, pareciera que están escalados. esto afecta a la luminosidad medida, PREGUNTAR. 

\begin{table}[] \centering \scriptsize

\caption{Emission line properties}
    \begin{tabular}{cccc}
    \hline \hline
    & SDSS  & NTT & SOAR \\
    \hline
    $W$ OIII$\lambda3133$ & - & 8.6$\pm$0.7 & -  \\
    FWHM  OIII$\lambda3133$ & - & 2509$\pm$16 & - \\
    
    $W$ H$\beta$ & 8.4$\pm$1.0 & 41.2$\pm$2.4 & 52.5$\pm$2.9  \\
    FWHM  H$\beta$ & 3485$\pm$102 & 2174$\pm$65 & 2315$\pm$41 \\
    $W$ NIII$\lambda4640$ & - & 9.20$\pm$1.3 & -\\
    FWHM NIII$\lambda4640$ & - & 2408$\pm$117  & -\\
    $W$ HeII$\lambda4685$ & 5.0$\pm$1.0 & 13.8$\pm$1.4 & 6.4$\pm$0.4 \\
    FWHM  HeII$\lambda4685$ & 5094$\pm$795 & 2500$\pm$25 & 2495$\pm$26\\
    $W$ H$\alpha\_1$ & 36.1$\pm$4.1 & - & 209$\pm$16.6  \\
    $W$ H$\alpha\_2$ & - & - & 75.1$\pm$4.1  \\
    FWHM  H$\alpha_1$ & 2870$\pm$192 & - & 1923$\pm$19 \\
    FWHM  H$\alpha_2$ & - & - &6954.4$\pm$75 \\
    \hline
    $L_{\rm 5100}$ & 43.5$\pm$0.09 & $43.7\pm0.09$ & $43.5\pm0.09$
    \\
    $M_{\rm BH,5100}$ & $7.8\pm0.12$ & $7.7\pm0.13$ & $7.6\pm0.13$ \\
    $\lambda_{\rm Edd,5100}$ & $0.071\pm0.026$ & $0.12\pm0.046$ & 0.1$\pm0.04$ \\
    $L_{\rm H\alpha}$ & $41.7\pm0.09$ & - & $42.5\pm0.09$\\
    $M_{\rm BH,H\alpha}$ & $7.4\pm0.05$& - & $7.4\pm0.06$\\
    \hline
    
    \end{tabular}\\
    \label{table:measurements}
    \medskip
    \justifying{{Notes: $W$ corresponds to the equivalent width in units of $\rm \AA$, while FWHM corresponds to the Full Width at Half Maximum in units of km s$^{-1}$}. Bolometric luminosity and black hole mass are expressed in units of erg s$^{-1}$ and M$_\odot$, respectively.  }
\end{table}

    We obtained the black hole mass through the classical relation $M\mathrm{_{BH}}=f\mathrm{_{BLR}} \cdot  R\mathrm{_{BLR}} \cdot { v}^2/G$, where $G$ is the gravitational constant, $f_{\rm BLR}$ is the virial factor\footnote{$f\mathrm{_{BLR}}=\left({{\mathrm{FWHM_{H\beta}}}}\,/\,{4550 {\pm 1000}}\right)^{ -1.17}$} \citep{mejia-restrepoetal18a}, $R_{\rm BLR}$ is the size of the broad line region obtained from the H$\beta$ Radius-Luminosity\footnote{$\mathrm{log}\left(\frac{R\mathrm{_{BLR}}}{\mathrm{1 lt-day}}\right)=(1.527\,\pm\,0.31)\, + \nonumber  \,0.533{^{+0.035}_{-0.033}}\, \mathrm{log}\left(\frac{ L_{\rm opt}}{10^{44}L_\odot}\right)$} \citep{bentz13}, and $v$ is the the velocity field of the broad line region represented by the FWHM of H$\beta$. The black hole mass for the observations is similar within uncertainties with a median value of 7.7 M$_\odot$. The Eddington ratio increases by a factor of 1.7 in the NTT and SOAR spectra with respect to the SDSS observation. This confirms the presence of a hard UV continuum in the post-flare observations. 
    
    The convolution between the accretion disk and host galaxy contributions could carry out uncertainties on the continuum flux and then in the black hole mass and Eddington ratio estimations. Alternatively, the black hole mass can be estimated with the luminosity and FWHM of the H$\alpha$ emission lines following the description of \citet{reines2013}. We obtained a black hole mass of 7.4$\pm$0.05 M$\odot$ and  7.4$\pm$0.06 M$_\odot$ for the SDSS and SOAR spectra. These values are 0.2-0.4 dex compared with the black hole mass estimated from continuum luminosity.
    
    The source is cataloged as a Narrow Line Seyfert 1 (NLS1) according to the flux ratio [OIII $\lambda$5007]/H$\beta$=0.9. In a BPT diagram, the target is located in the transition region between starforming and LINERs. This means that the ionizating radiation comes from the AGN plus young stars. 
    
%    F(HeII)+F(NIII)/Hb = 0.63 dentro de los rangos que reportan para las BBF
    
 %    f_fex_6375   113.685890       106.046806 
    
    \subsection{Neil Gehrels Swift Observatory}
    
    The {\it Swift} X-ray Telescope \citep[XRT,][]{Burrows05}, onboard the Neil Gehrels Swift Observatory, observed AT2023zgo in the Photon Counting (PC) between January 17th and May 29th, 2024, approximately once per week, taking a total of 18 observations. 
    The source was not detected on individual observations, and count rates and upper limits were retrieved from the living Swift-XRT point source catalogue  \citep{LSXPSCatalogue2023}, and converted into flux using WebPIMMS and a power law model using a spectral index of 2 and galactic absorption 2.62$\times$10$^{20}$cm$^{-2}$.
    The flux upper limits are in the range between 1.2$\times$10$^{-13}$ and 5.1$\times$10$^{-13}$ erg s$^{-1}$cm$^{-2}$.
    Stacking all the observations results in a 24 ksec observation and a 2$\sigma$ detection with a flux of 1.1[0.4-1.7]$\times$10$^{-14}$ erg s$^{-1}$cm$^{-2}$.
    \begin{figure}
    \includegraphics[width=8cm]{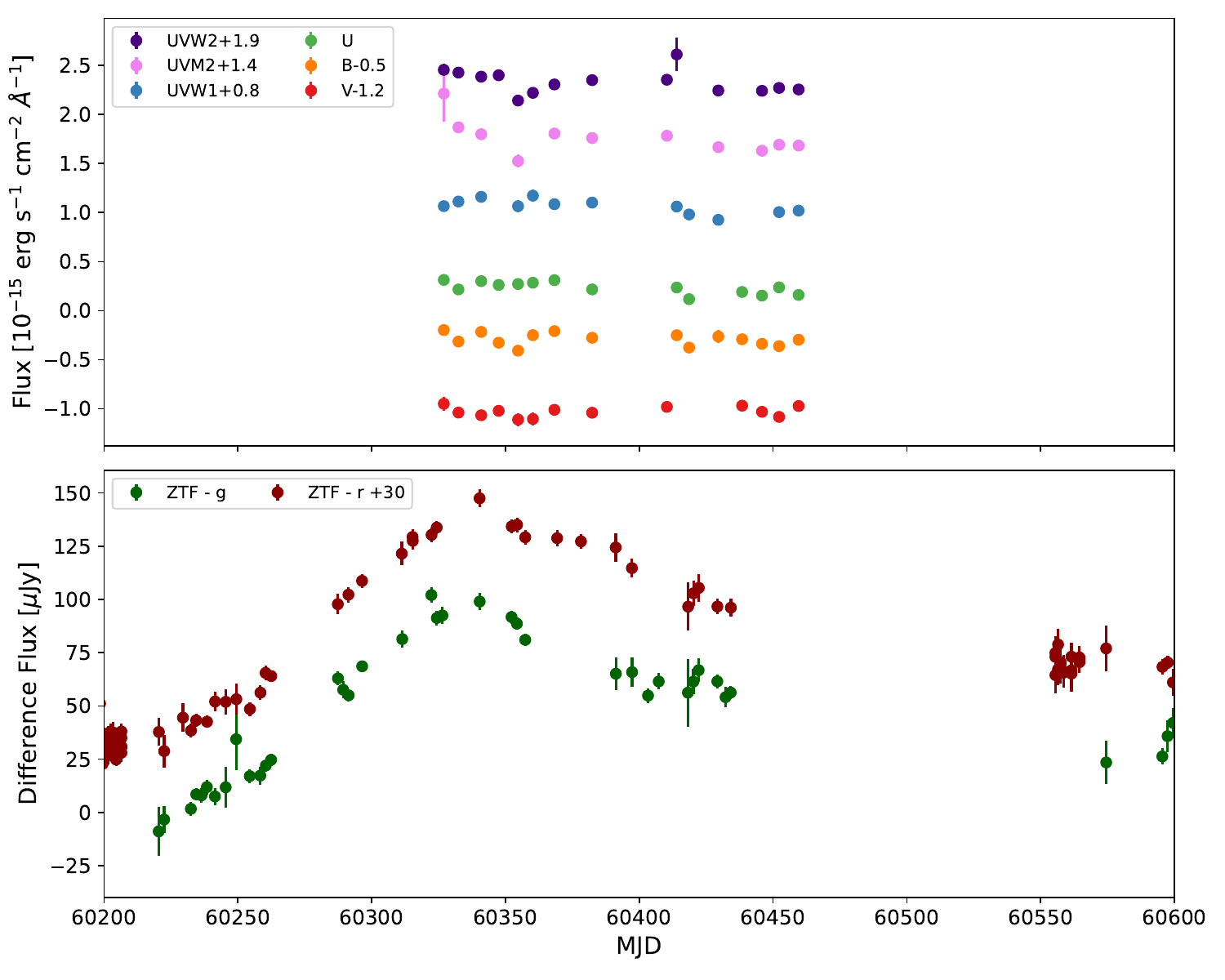}
    \centering
    \caption{Swift/UVOT (in total flux) and ZTF forced photometry (in difference flux) data between September 13, 2023, and October 17, 2024.}
    \label{fig:swiftztfflux}
    \end{figure}
   
    The Ultraviolet and Optical Telescope  \citep[UVOT,][]{2005SSRv..120...95R}
    observes simultaneously with the X-ray observations on each epoch.
    Observations were obtained with the six primary photometric filters: V (centred at 5468 \AA), B (at 4392 \AA), U (at 3465 \AA), UVW1 (at 2600 \AA), UVM2 (at 2246 \AA) and UVW2 (at 1928 \AA). The {\sc uvotsource} task within software HEASoft version 6.33 was used to perform aperture photometry using a circular aperture of radius 5 arcsec centred on the coordinates of AT2023zgo. The background region was selected free of sources adopting a circular region of 20 arcsec close to the nucleus. 
    There are certain areas in the UVOT detector where the throughput is lower than for the rest of the detector, so observations affected by this effect were not taken into account. 
    
\begin{figure}
    \includegraphics[width=8
    cm]{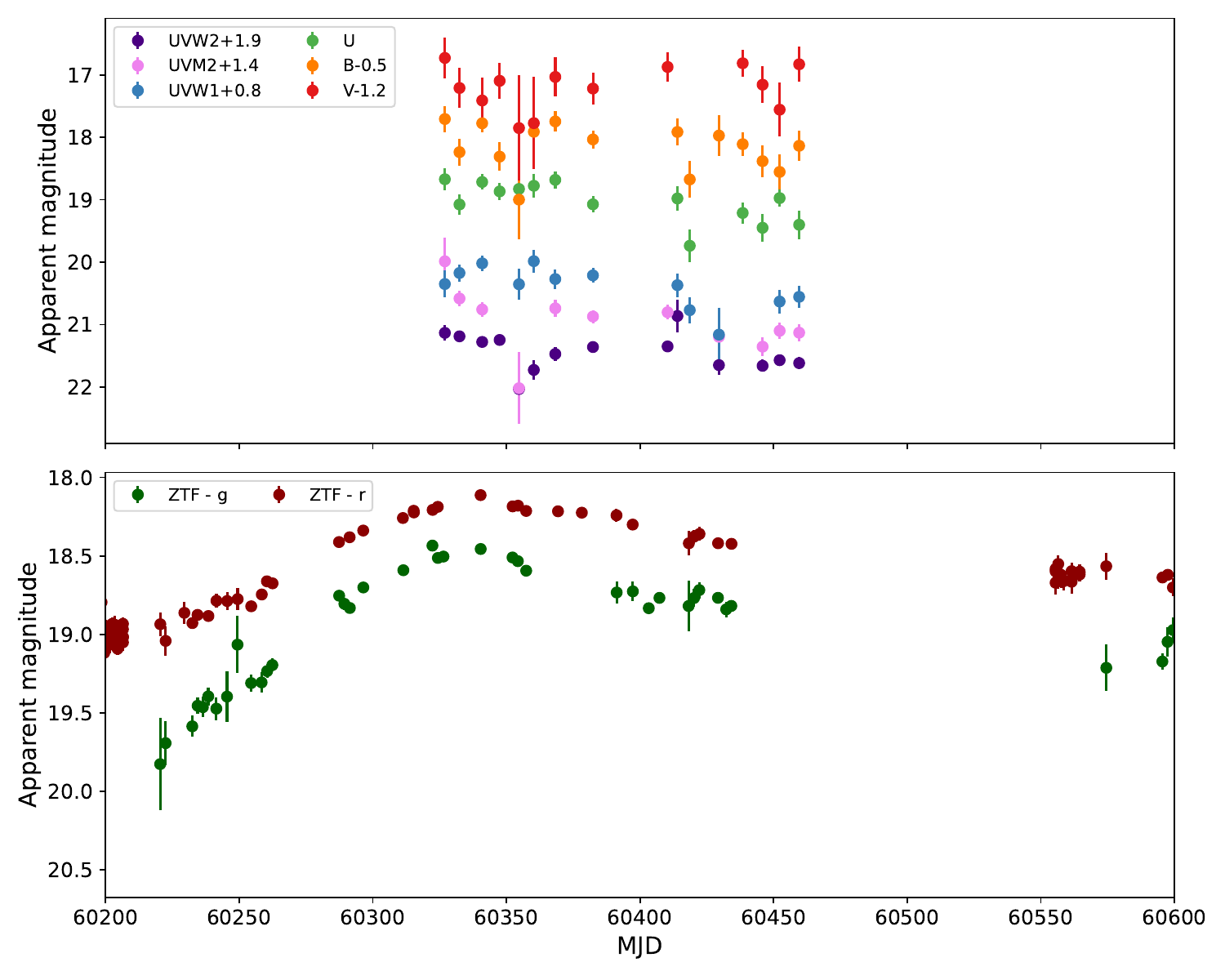}
    \centering
    \caption{Swift/UVOT and ZTF forced photometry data in apparent magnitudes between September 13, 2023, and October 17, 2024.}
    \label{fig:swiftztfmag}
    \end{figure}

    The ZTF light curve of AT2023zgo is presented in difference flux (i.e., the variable flux) in Fig. \ref{fig:swiftztfflux}, with the {\it Swift}/UVOT measurements in total flux, and in apparent magnitude (i.e., the magnitude corrected for the contribution of the host galaxy as measured in the reference image) in Fig. \ref{fig:swiftztfmag}, with the {\it Swift}/UVOT measurements in apparent magnitude. It is worth noting a very similar shape in this plot in the optical and UV data.

  \newpage  
\section{AGN Spectra of Counterpart Candidates}
Here we present the archival spectra before the onset of flare. This was essentially used for calculating the SMBH masses.
\begin{figure}[h]
    \centering
    % --- First row ---
    \begin{subfigure}[b]{0.6\textwidth}
        \includegraphics[width=\textwidth]{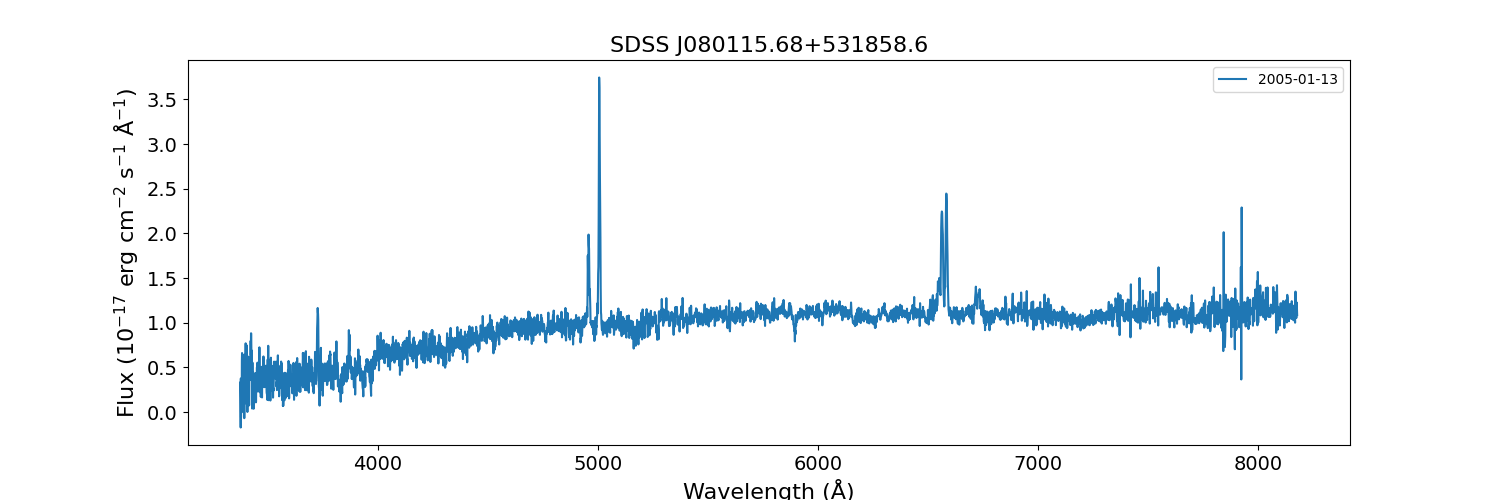}
        \caption{ZTF24absmrlr}
    \end{subfigure}
    \hfill
    \begin{subfigure}[b]{0.6\textwidth}
        \includegraphics[width=\textwidth]{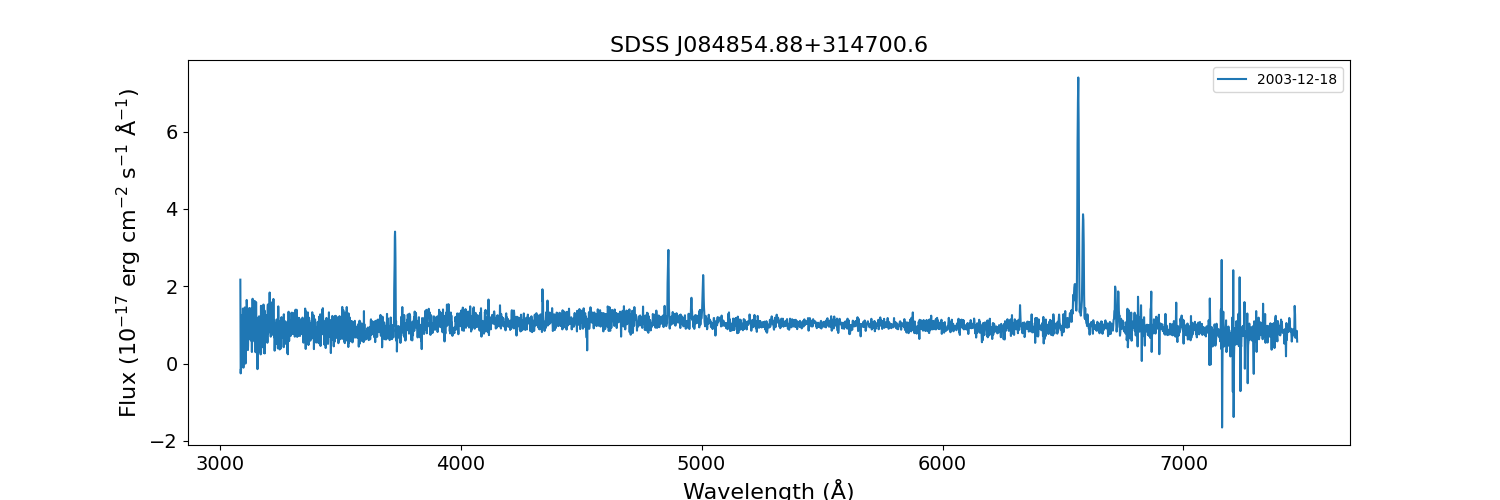}
        \caption{AT2023zgo/ ZTF23abqkwzr}
    \end{subfigure}

    % --- Second row ---
    \vskip\baselineskip
    \begin{subfigure}[b]{0.6\textwidth}
        \includegraphics[width=\textwidth]{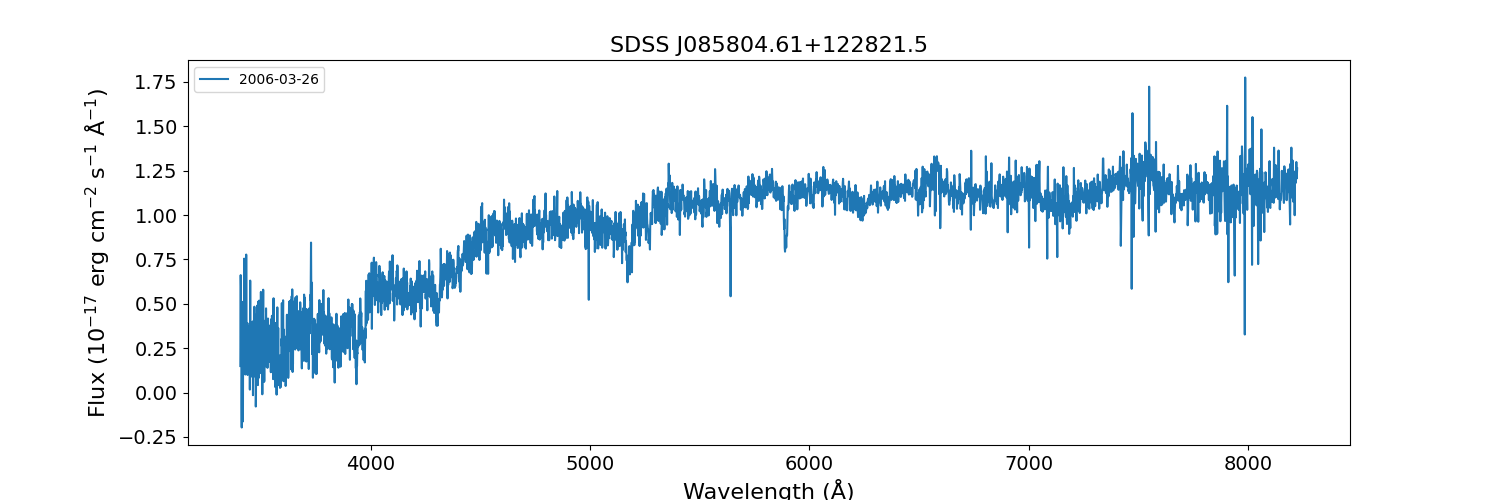}
        \caption{ZTF24abricne}
    \end{subfigure}
    \hfill
    \begin{subfigure}[b]{0.6\textwidth}
        \includegraphics[width=\textwidth]{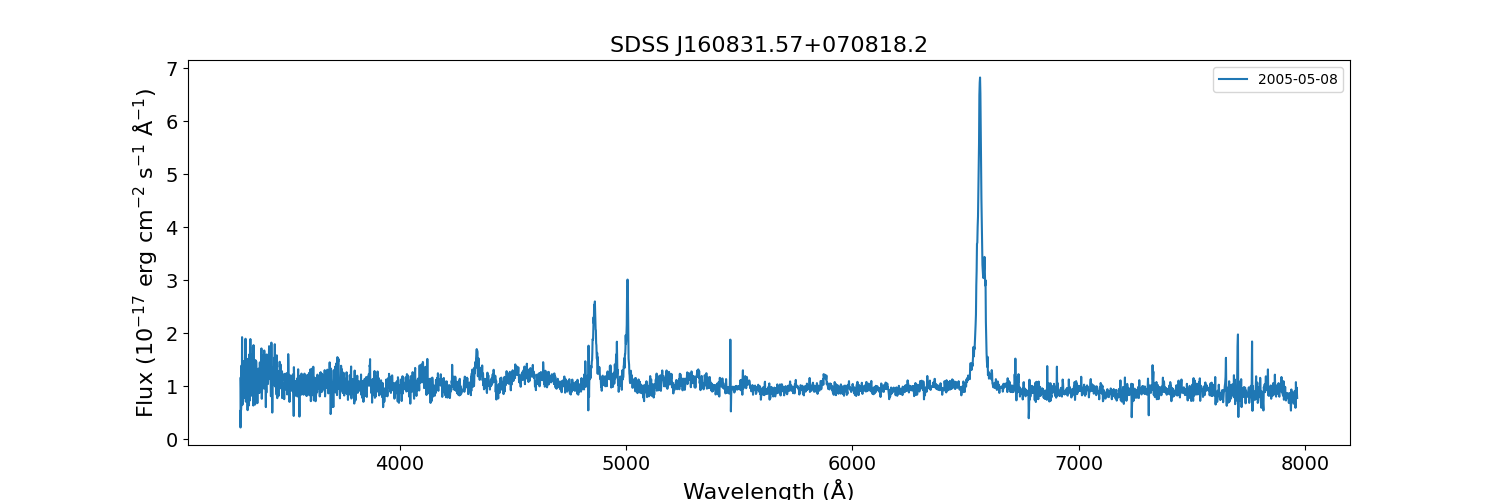}
        \caption{AT2025cze / ZTF25aafwffi}
    \end{subfigure}

    \caption{SDSS Archival spectra of the host AGNs for the counterpart candidates}
    \label{fig:fourplots}
\end{figure}  

\end{appendix}
\label{LastPage}
\end{document}